\documentclass[
  aps,
  prd,
  preprint,
  preprintnumbers,
  nofootinbib,
  nobibnotes,
]{revtex4-1}

\pdfoutput=1

\usepackage{amssymb,amsmath}
\usepackage{color}
\usepackage{graphicx,subfigure}
\usepackage{hyperref}
\usepackage{float} 
\usepackage[T1]{fontenc}
\usepackage{braket}

\newcommand{\nn}{\nonumber}

\begin{document}

\title{Enhancement of charm CP violation due to nearby resonances}

\author{Stefan Schacht}
\email{stefan.schacht@manchester.ac.uk}
\affiliation{Department of Physics and Astronomy, University of Manchester, Manchester M13 9PL, United Kingdom}
\author{Amarjit Soni}
\email{adlersoni@gmail.com}
\affiliation{Physics Department, Brookhaven National Laboratory, Upton, NY 11973, USA}

\begin{abstract}
Quantitative understanding of CP violation is extremely important as naturalness reasoning strongly suggests that new physics should be accompanied by beyond the Standard Model CP-odd phases. In 2019 LHCb made the first $5 \sigma$ discovery of CP violation in the charm system, leading to the new world average $\Delta a_{CP}^{\mathrm{dir}} = -0.00161 \pm 0.00028$. While some calculations have found this observation as requiring new physics, we suggest that scalar resonances nearby to $m_{D^0}$, in particular, $f_0(1710)$ and/or $f_0(1790)$ cause enhancements for CP violation within the SM. Thereby, our calculations based on the Standard Model suggest compatibility with the LHCb observations for now. However, experimental information especially on $f_0(1790)$ is rather sparse limiting our accuracy and further data on these resonances is strongly advocated.
\end{abstract}

\maketitle

\section{Introduction and Motivation \label{sec:intro}}

Naturalness reasoning strongly suggests that beyond the Standard Model (BSM) CP-odd phases should exist.  
The 1964 experiment at BNL by Christenson et al~\cite{Christenson:1964fg} showed for the first time that CP is not a 
symmetry of nature. Therefore, any interaction that we write down that can be complex, allowing for CP violation to occur,  
simply has to be complex since there is no symmetry that requires it to be real. 

But to discern the effects of 
new physics (NP) from that of the Standard Model (SM) requires precision both in experimental measurements as well as in 
theoretical predictions from the SM. More often than not the latter task can be very challenging because of the 
influence of non-perturbative effects of QCD on the relevant channels for $K$, $D$ or $B$ decays involved in direct CP 
violation studies. 
This explicitly manifests itself in the well-known  examples of $\epsilon'$ for 
$K \to \pi \pi$~\cite{Abbott:2020hxn, Aebischer:2020jto, Cirigliano:2019ani},
$A_{CP}$ measurements in the four $B \to K, \pi$ modes~\cite{Zyla:2020zbs,Aaij:2020wnj}, and the more relevant example to this work, $\Delta A_{CP}$ in $D^0$ decays, where~\cite{Amhis:2019ckw, Aaij:2019kcg, BaBar:2007tfw, CDF:2012ous, LHCb:2014kcb, LHCb:2016csn, LHCb:2019hro}
\begin{align}
\Delta A_{CP} \approx \Delta a_{CP}^{\mathrm{dir}} &\equiv a_{CP}^{\mathrm{dir}}(D^0\rightarrow K^+K^-) - a_{CP}^{\mathrm{dir}}(D^0\rightarrow \pi^+\pi^-) \label{eq:DeltaACP}\\
&=-0.00161 \pm 0.00028\,,
\end{align}
and
\begin{align}
a_{CP}^{\mathrm{dir}}(f) &\equiv \frac{
|\mathcal{A}(D^0\rightarrow f) |^2 - |\mathcal{A}(\overline{D}^0\rightarrow f)|^2  
}{ 
|\mathcal{A}(D^0\rightarrow f) |^2 + |\mathcal{A}(\overline{D}^0\rightarrow f)|^2  
}\,.
\end{align}
Note that for charm decays to CP eigenstates the indirect contributions to CP violation $a_{CP}^{\mathrm{ind}}$  are universal to a very good approximation, and we can write the time-integrated CP asymmetry as~\cite{Grossman:2006jg}
\begin{align}
A_{CP} &\approx a_{CP}^{\mathrm{dir}} + a_{CP}^{\mathrm{ind}}\,, \label{eq:indirect} \\ 
a_{CP}^{\mathrm{ind}} &\equiv -\eta_f^{CP}\frac{y}{2} \left( \left|\frac{q}{p}\right| - \left|\frac{p}{q}\right| \right) \cos\phi +
		      \eta_f^{CP}\frac{x}{2} \left( \left|\frac{q}{p}\right| + \left|\frac{p}{q}\right| \right) \sin\phi\,,
\end{align}
with the weak mixing phase $\phi$ and using the standard notation for the mixing parameters~\cite{Grossman:2006jg}.
It follows that $a_{CP}^{\mathrm{ind}}$ cancels in the difference of CP asymmetries in Eq.~(\ref{eq:DeltaACP}).

Recently, LHCb and Belle also improved the precision of the measurement of the CP asymmetry of charm decays into neutral kaons, leading to the new average~\cite{Amhis:2019ckw, LHCb:2021rdn, Dash:2017heu, LHCb:2015ope, CLEO:2000opx}
\begin{align}
a_{CP}^{\mathrm{dir}}(D^0\rightarrow K_SK_S) &= -0.019\pm 0.010\,,
\end{align}
bringing the measurement closer to a possible second observation of CP violation in the charm system~\cite{Nierste:2015zra}.

For the case at hand of $\Delta A_{CP}$, there are theoretical frameworks~\cite{Chala:2019fdb, Khodjamirian:2017zdu} which seem to suggest that the SM Cabibbo-Kobayashi-Maskawa (CKM) 
paradigm cannot account for the value measured by LHCb whereas others, such as ours, seem to model non-perturbative 
effects in such a way that no significant difference exists at present between the experimental measurements and their estimated value based on the SM. 
After the measurement presented in Ref.~\cite{Aaij:2019kcg}, different theoretical interpretations have been brought forward~\cite{Grossman:2019xcj,Chala:2019fdb, Li:2019hho, Soni:2019xko, Cheng:2019ggx, Buras:2021rdg, Acaroglu:2021qae} (see for earlier works Refs.~\cite{Einhorn:1975fw, Abbott:1979fw,Golden:1989qx,  Brod:2012ud, Bhattacharya:2012ah, Franco:2012ck, Hiller:2012xm, Nierste:2017cua, Nierste:2015zra, Muller:2015rna, Grossman:2018ptn, Buccella:1994nf, Grossman:2006jg, Artuso:2008vf, Khodjamirian:2017zdu, Buccella:2013tya, Cheng:2012wr, Feldmann:2012js, Li:2012cfa, Atwood:2012ac, Grossman:2012ry, Buccella:2019kpn, Yu:2017oky, Brod:2011re,Pirtskhalava:2011va}). It is an open question if the observed amount of CP violation is to be interpreted as new physics or not.

Making precise theoretical predictions for the case of charm decays is particularly challenging because of the presence of many nearby resonances. These require a reliable non-perturbative framework. In recent  years lattice methods have made some progress in this regard and are starting to show 
limited success~\cite{Johnson:2020ilc, Rendon:2020rtw, Fischer:2020fvl, Bruno:2020kyl}, 
but are not quite ready yet to tackle $\Delta A_{CP}$ for charm decays. Therefore, we have to resort to phenomenological models.

As is well known, singly-Cabibbo suppressed charm decays are dominated by tree operators. 
Through rescattering, penguin contractions of tree operators are formed, which have a relative weak and strong phase with respect 
to the tree amplitude, leading to a non-vanishing CP asymmetry~\cite{Brod:2012ud, Grossman:2019xcj}. 
An exception is the decay $D^0\rightarrow K_SK_S$, where the dominant contribution to CP violation comes from the interference of SU(3)$_F$-breaking and SU(3)$_F$-leading exchange diagrams,~\emph{i.e.},~CP violation is even present for 
vanishing penguin annihilation~\cite{Nierste:2015zra, Nierste:2017cua, Hiller:2012xm, Atwood:2012ac}.

In this paper, we work out a specific dynamical mechanism of charm CP violation, namely the possible enhancement of rescattering through scalar resonances close to the $D^0$ mass. The possible role of the scalar resonances for charm CP violation has recently been pointed out in Ref.~\cite{Soni:2019xko}. In Ref.~\cite{Soni:2019xko} only the impact of the $f_0(1710)$ resonance has been studied. Here, we study the implications of the picture of Ref.~\cite{Soni:2019xko} in more detail and take also the implications of the $f_0(1790)$ resonance into account, which is even closer to $m_{D^0}$.\footnote{Thus this work supersedes Ref.~\cite{Soni:2019xko}.} Note that the $f_0(1790)$ is not yet fully established and so far not listed in Ref.~\cite{Zyla:2020zbs}.
It has first been seen by BESII~\cite{Ablikim:2004wn}, and has also been seen by LHCb~\cite{Aaij:2014emv}.\footnote{Note however, that the $f_0(1790)$ resonance has not been seen in the analysis of CLEO data in Ref.~\cite{Dobbs:2015dwa}.} 

As the masses of these resonances are close to $m_{D^0}$, their effect might be significant. It is therefore reasonable to investigate their role for CP violation further. Unfortunately, available experimental information on the $f_0(1790)$ resonance is rather sparse and this will limit our ability to be precise.

The role of resonances for nonleptonic charm decays has been studied 
for a long time~\cite{Donoghue:1979fp, Lipkin:1980es, Sorensen:1981vu, Kamal:1988ub, Buccella:1990sp, Buccella:1992ym, Buccella:1996uy, Fajfer:1999hh, Rosner:1999xd, Gronau:1999zt, Cheng:2010ry, Fusheng:2011tw, Biswas:2015aaa, Gronau:2015zga}, including also corresponding implications for 
CP violation \cite{Eilam:1991yv, Atwood:1994zm, Li:2012cfa, Li:2013xsa, Buccella:1992sg, Buccella:2013tya, Buccella:2019kpn, Dery:2021mll}. The possible impact of scalar resonances on charm CP violation has also been mentioned in Ref.~\cite{Khodjamirian:2017zdu}.
Resonances play a role in the discussion of 
charm mixing, too~\cite{Falk:1999ts,Golowich:1998pz, Bergmann:2000id, Falk:2001hx, Falk:2004wg}.

Note that with CP asymmetries measured in more channels in the future, the CPT theorem might have important implications for 
the interplay of CP violation with final state interactions between different decay channels that share the same quantum numbers, i.e.~which can rescatter into each other, see Refs.~\cite{Gerard:1988jj, Wolfenstein:1990ks, Gerard:1990ni, Atwood:1997iw, Atwood:2000tu, Atwood:2012ac, Zwicky:2007vv, Bediaga:2020qxg}.

\section{Isospin decomposition \label{sec:isospin}}

Employing the notation of Refs.~\cite{Muller:2015lua, Muller:2015rna}, we write the amplitudes of 
singly-Cabibbo suppressed (SCS) decays as 
\begin{align}
\mathcal{A} &= \lambda_{sd} A_{sd} - \frac{\lambda_b}{2} A_b\,, \label{eq:general-decomposition}
\end{align}
where we use the CKM matrix element combinations
\begin{align}
\lambda_{sd} &= \frac{\lambda_s-\lambda_d}{2}\,, \quad 
\lambda_q     = V_{cq}^* V_{uq}\,. 
\end{align}
The amplitudes are related to the branching ratio as 
\begin{align}
\mathcal{B}(D\rightarrow P_1P_2) &= \vert \mathcal{A}\vert^2 \, \mathcal{P}\,, 
\end{align}
with the phase space factor
\begin{align}
\mathcal{P} &=  \frac{\tau_D}{16\pi m_D^3} \sqrt{(m_D^2 - (m_{P_1} - m_{P_2})^2 )  ( m_D^2 - (m_{P_1} + m_{P_2})^2)}\,.
\end{align}
Note that the phase space has mass dimension $[\text{GeV}^{-2}]$, such that the amplitudes have mass dimension $[\text{GeV}]$. 

In this notation the direct CP asymmetries of SCS decays are given as 
\begin{align}
a_{CP}^{\mathrm{dir}} &= \frac{\vert\mathcal{A}\vert^2  - \vert \overline{\mathcal{A}}\vert^2 
				}{  
				\vert\mathcal{A}\vert^2  + \vert \overline{\mathcal{A}}\vert^2
				}
	= \mathrm{Im}\left(\frac{\lambda_b}{\lambda_{sd}}\right) \mathrm{Im}\left( \frac{A_b}{A_{sd}}\right)\,,
\end{align}
where~\cite{Zyla:2020zbs}  
\begin{align}
\mathrm{Im}\left(\frac{\lambda_b}{\lambda_{sd}}\right) &= -0.00059\pm 0.00003\,.
\end{align}
In the following, we consider the impact of resonances with definite isospin quantum numbers. Therefore, we parametrize the SCS charm decays in terms of isospin matrix elements. 
We adapt the notation from Refs.~\cite{Atwood:2012ac, Franco:2012ck} in such a way as to make the dependence on the CKM matrix elements explicit. In this way, the hadronic matrix elements have no absorbed CKM matrix elements and carry a strong phase only: 
\begin{align}
\mathcal{A}(D^+\rightarrow \pi^+\pi^0) &= \frac{\sqrt{3}}{2} \lambda_{sd} A^{\pi\pi}_{\frac{3}{2}, 2}\,, \label{eq:isospin-first} \\
\mathcal{A}(D^0\rightarrow \pi^+\pi^-) &= \frac{1}{\sqrt{6}} \lambda_{sd} A_{\frac{3}{2},2}^{\pi\pi} + 
				\frac{1}{\sqrt{3}} \left( \lambda_{sd} A^{\pi\pi}_{\frac{1}{2},0} -\frac{\lambda_b}{2} B^{\pi\pi}_{\frac{1}{2},0}  \right) \,, \\
\mathcal{A}(D^0\rightarrow \pi^0\pi^0) &= \frac{1}{\sqrt{3}} \lambda_{sd} A_{\frac{3}{2},2}^{\pi\pi} - 
				\frac{1}{\sqrt{6}} \left( \lambda_{sd} A_{\frac{1}{2},0}^{\pi\pi} - \frac{\lambda_b}{2} B_{\frac{1}{2},0}^{\pi\pi} \right) \,,
\end{align}
and 
\begin{align}
-\sqrt{2} \mathcal{A}(D^+\rightarrow K^+K_S) &= -\frac{1}{2} \lambda_{sd} A_{\frac{3}{2},1}^{KK} + 
			    \left( \lambda_{sd} A_{\frac{1}{2},1}^{KK} -\frac{\lambda_b}{2} B_{\frac{1}{2},1}^{KK}  \right)\,, \\
\mathcal{A}(D^0\rightarrow K^+K^-) &= \frac{1}{2} \lambda_{sd} A_{\frac{3}{2},1}^{KK} + 
			    \frac{1}{2} \left( \lambda_{sd} A_{\frac{1}{2},1}^{KK} - \frac{\lambda_b}{2} B_{\frac{1}{2},1}^{KK} \right) + 
			    \frac{1}{2} \left( \lambda_{sd} A_{\frac{1}{2},0}^{KK} - \frac{\lambda_b}{2} B_{\frac{1}{2},0}^{KK}\right)\,,\\
-\sqrt{2} \mathcal{A}(D^0\rightarrow K_SK_S) &= \frac{1}{2} \lambda_{sd} A_{\frac{3}{2},1}^{KK} + 
			    \frac{1}{2} \left( \lambda_{sd} A_{\frac{1}{2},1}^{KK} - \frac{\lambda_b}{2} B_{\frac{1}{2},1}^{KK}  \right) - 
			    \frac{1}{2} \left( \lambda_{sd} A_{\frac{1}{2},0}^{KK} - \frac{\lambda_b}{2} B_{\frac{1}{2},0}^{KK}  \right)\,. \label{eq:isospin-last}
\end{align}
Here, the superscripts of the hadronic matrix elements indicate the final state, the first subscript corresponds to the isospin of the operator in the Hamiltonian, and the second subscript indicates the isospin of the final state.
Note that the contribution of the $\Delta I=3/2$ Hamiltonian to the CKM-subleading amplitude $A_b$ is negligible in the SM~\cite{Grossman:2012eb, Franco:2012ck}. 

Working in the isospin limit, we have \cite{Buccella:1992sg,Grossman:2012eb}
\begin{align}
a_{CP}^{\mathrm{dir}}(D^0\rightarrow \pi^+\pi^0) &= 0\,, \label{eq:pipi-isospin-1}
\end{align}
and the sum rule~\cite{Kwong:1993ri, Grossman:2012eb,Atwood:2012ac}
\begin{align}
- \mathcal{A}(D^+\rightarrow \pi^+\pi^0) + 
\frac{1}{\sqrt{2}}  \mathcal{A}(D^0\rightarrow \pi^+\pi^-) + 
\mathcal{A}(D^0\rightarrow \pi^0\pi^0) = 0\,. \label{eq:pipi-isospin-2}
\end{align}
For the $D\rightarrow KK$ system there are no isospin relations analogous to Eqs.~(\ref{eq:pipi-isospin-1}) and (\ref{eq:pipi-isospin-2}).

As we are only sensitive to relative phases, we choose $A_{\frac{3}{2},2}^{\pi\pi}$ to be real and positive.
We employ therefore as real theory parameters of $D\rightarrow \pi\pi$: 
\begin{align}
A_{\frac{3}{2},2}^{\pi\pi}\,,
\vert A_{\frac{1}{2},0}^{\pi\pi}\vert\,,
\mathrm{arg}\left( A_{\frac{1}{2},0}^{\pi\pi}\right), 
\vert B_{\frac{1}{2},0}^{\pi\pi}\vert\,,
\mathrm{arg}\left( B_{\frac{1}{2},0}^{\pi\pi}\right).
\end{align}
Regarding the $D\rightarrow KK$ decays, we choose $A_{\frac{3}{2},1}^{KK}$ to be real and positive and employ the 
real theory parameters 
\begin{align}
 A_{\frac{3}{2},1}^{KK},\,
\vert A_{\frac{1}{2},1}^{KK}\vert,\,
\vert A_{\frac{1}{2},0}^{KK}\vert,\,
\mathrm{arg} \left( A_{\frac{1}{2},1}^{KK}\right),\,
\mathrm{arg} \left( A_{\frac{1}{2},0}^{KK}\right),\, \nn\\
 \vert B_{\frac{1}{2},1}^{KK}\vert,\,
\vert B_{\frac{1}{2},0}^{KK}\vert, 
\mathrm{arg} \left( B_{\frac{1}{2},1}^{KK}\right),\,
\mathrm{arg} \left( B_{\frac{1}{2},0}^{KK}\right).
\end{align}
In our numerical analysis, we vary all strong phases in their complete range.

\section{Resonance Model \label{sec:resonance}} 

On top of the isospin decomposition Eqs.~(\ref{eq:isospin-first})--(\ref{eq:isospin-last}), we make the following assumptions:
\begin{itemize}
\item The decay amplitudes to $I=0$ states 
$A_{\frac{1}{2},0}^{\pi\pi}$, $B_{\frac{1}{2},0}^{\pi\pi}$, $A_{\frac{1}{2},0}^{KK}$, $B_{\frac{1}{2},0}^{KK}$ are dominated by scalar resonances, namely $f_0(1710)$ and/or $f_0(1790)$, 
with quantum numbers $I=0$ and $J^{PC} = 0^{++}$.
\item The $I=1$ amplitudes are relatively suppressed:
\begin{align}
\left|\frac{A_{\frac{1}{2},1}^K}{A_{\frac{1}{2},0}^K}\right| &\lesssim 40\%\,, &
\left|\frac{A_{\frac{3}{2},1}^K}{A_{\frac{1}{2},0}^K}\right| &\lesssim 40\%\,, &
\left|\frac{B_{\frac{1}{2},1}^K}{B_{\frac{1}{2},0}^K}\right| &\lesssim 40\%\,, \label{eq:iso1-suppression}
\end{align}
reflecting the dominance of the scalar resonances. 
\end{itemize}
In our numerical study we implement Eq.~(\ref{eq:iso1-suppression}) with a Gaussian error.
With more data, in the future the assumptions of our resonance model can be revisited and improved.

Our ansatz employs the strong effective couplings $g_{f_0PP'}$ of the resonance and the pseudoscalars, which we define through the equation  
\begin{align}
\mathcal{B}(f_0\rightarrow PP') &\equiv g_{f_0 PP'}^2 \mathcal{P}(f_0\rightarrow PP')\,. 
\end{align}
Furthermore, we define the hadronic matrix elements
\begin{align}
M_{f_0}^{sd} &\equiv \bra{f_0} \mathcal{O}^{\Delta I=1/2}_{sd}\ket{D^0}\,, \qquad 
M_{f_0}^{b}  \equiv \bra{f_0} \mathcal{O}^{\Delta I=1/2}_{b}\ket{D^0}\,, \label{eq:hadronic-MEs} 
\end{align} 
where $\mathcal{O}_{sd}^{\Delta I=1/2}$ is the $\Delta I=1/2$ operator contributing to $A_{sd}$ and  
$\mathcal{O}_{b}^{\Delta I=1/2}$ is the $\Delta I=1/2$ operator contributing to $A_{b}$ in Eq.~(\ref{eq:general-decomposition}). 

Due to the lack of precise data on the scalar resonances, including the hadronic matrix elements 
Eq.~(\ref{eq:hadronic-MEs}), we consider three simplified scenarios:
\begin{enumerate}
\item[(1)] Dominance of $f_0(1710)$ only. 
\item[(2)] Dominance of $f_0(1790)$ only. 
\item[(3)] Contribution of both $f_0(1710)$ and $f_0(1790)$, but with the same hadronic matrix elements for both scalar $f_0$ resonances $M_{f_0}^{sd}$ and $M_{f_0}^{b}$. 
\end{enumerate}
Our implementation of these scenarios does not depend on the details of the functional shape of the resonance 
amplitude $R( m_{f_0}, \Gamma_{f_0}, m_{D^0})$. Specific examples can be found in
Refs.~\cite{Choi:1998yx,Fajfer:1999hh, Cheng:2002wu, Li:2019hho, Soni:2019xko, Li:2012cfa, Gronau:1999zt, Colangelo:2010bg, Ropertz:2018stk,Kou:2018nap}.
For scenario (1) and (2) our simplified resonance model ansatz reads: 
\begin{align}
A_{\frac{1}{2},0}^{\pi\pi} &= g_{f_0\pi\pi} M_{f_0}^{sd} R( m_{f_0}, \Gamma_{f_0}, m_{D^0})\,,  \label{eq:res-model-1} \\
B_{\frac{1}{2},0}^{\pi\pi} &= g_{f_0\pi\pi} M_{f_0}^{b} R( m_{f_0}, \Gamma_{f_0}, m_{D^0})\,,  \label{eq:res-model-2} \\
A_{\frac{1}{2},0}^{KK}     &= g_{f_0KK} M_{f_0}^{sd}R( m_{f_0}, \Gamma_{f_0}, m_{D^0})\,,  \label{eq:res-model-3}\\
B_{\frac{1}{2},0}^{KK}     &= g_{f_0KK} M_{f_0}^{b}R( m_{f_0}, \Gamma_{f_0}, m_{D^0})\,. \label{eq:res-model-4}
\end{align}
In Eqs.~(\ref{eq:res-model-1})--(\ref{eq:res-model-4}), $f_0$ is either $f_0(1710)$ or $f_0(1790)$.
In scenario (3), we have 
\begin{align}
A_{\frac{1}{2},0}^{\pi\pi} &= M_{f_0}^{sd} \, R( g_{f_0(1710)\pi\pi}, g_{f_0(1790)\pi\pi}, m_{f_0(1710)}, \Gamma_{f_0(1710)},  m_{f_0(1790)}, \Gamma_{f_0(1790)},  m_{D^0} )\,, \label{eq:res-model-5} \\
B_{\frac{1}{2},0}^{\pi\pi} &= M_{f_0}^{b} \, R( g_{f_0(1710)\pi\pi}, g_{f_0(1790)\pi\pi}, m_{f_0(1710)}, \Gamma_{f_0(1710)},  m_{f_0(1790)}, \Gamma_{f_0(1790)},  m_{D^0} ) \,,  \label{eq:res-model-6} \\
A_{\frac{1}{2},0}^{KK}    &=  M_{f_0}^{sd}\, R( g_{f_0(1710)KK}, g_{f_0(1790)KK}, m_{f_0(1710)}, \Gamma_{f_0(1710)},  m_{f_0(1790)}, \Gamma_{f_0(1790)},  m_{D^0} ) \,,  \label{eq:res-model-7}\\
B_{\frac{1}{2},0}^{KK}    &=  M_{f_0}^{b}\, R( g_{f_0(1710)KK}, g_{f_0(1790)KK}, m_{f_0(1710)}, \Gamma_{f_0(1710)},  m_{f_0(1790)}, \Gamma_{f_0(1790)},  m_{D^0} )\,, \label{eq:res-model-8}
\end{align}
where the $R$-function denotes now the unknown interference of the $f_0(1710)$ and $f_0(1790)$ resonances including 
possible mixing effects. The data in Table~\ref{tab:resonances} indicates that the resonances $f_0(1710)$ and $f_0(1790)$ are 
overlapping and therefore, in general, mixing between the scalar resonances takes place. 

In all three simplified scenarios, we express four isospin amplitudes by two hadronic matrix elements, which are  
defined in Eq.~(\ref{eq:hadronic-MEs}). It follows 
\begin{align}
\frac{ A_{\frac{1}{2},0}^{\pi\pi} }{ A_{\frac{1}{2},0}^{KK} } &= 
\frac{ B_{\frac{1}{2},0}^{\pi\pi} }{ B_{\frac{1}{2},0}^{KK} } \equiv r_{f_0}\,, \label{eq:amplitude-ratio}  
\end{align}
where $r_{f_0}$ does not depend anymore on $M_{f_0}^{sd}$ or $M_{f_0}^b$. 

In scenarios (1) and (2), we have 
\begin{align}
\vert r_{f_0}\vert  &=  \left|\frac{g_{f_0\pi\pi} }{ g_{f_0KK} }\right|  = \sqrt{
		\frac{\mathcal{B}(  f_0 \rightarrow \pi\pi   )}{\mathcal{B}( f_0\rightarrow KK )}
		\frac{\mathcal{P}( f_0 \rightarrow KK )}{\mathcal{P}( f_0\rightarrow \pi\pi  )} 
	}\,. \label{eq:resonance-ratio}
\end{align}
Only in scenario (1) we have experimental information on $\vert r_{f_0}\vert$, namely~\cite{Zyla:2020zbs}
\begin{align}
\frac{\mathcal{B}(f_0(1710)\rightarrow \pi\pi)}{\mathcal{B}(f_0(1710)\rightarrow KK)} &= 0.23\pm 0.05\,. \label{eq:ratio-1710}
\end{align}
Absolute branching ratios of $f_0(1710)$ or $f_0(1790)$ are not quoted by the PDG~\cite{Zyla:2020zbs}. 
Clearly, any further experimental information would improve our results considerably.

Due to lack of experimental information, in scenarios (2) and (3) we leave $\vert r_{f_0}\vert$ unconstrained and extract the corresponding value needed in order to explain the charm decay data from our fit. In scenario (2) this leads to a prediction for the ratios of branching ratios to pions and kaons, analogous to Eq.~(\ref{eq:ratio-1710}), see Sec.~\ref{sec:predictions} below. In scenario~(3), the result can be interpreted with more information on the scalar resonances in the future.

In all three considered scenarios, from Eq.~(\ref{eq:amplitude-ratio}) it follows furthermore the following relation between the relevant strong phases:
\begin{align}
\mathrm{arg}(A_{\frac{1}{2},0}^{\pi\pi}) - \mathrm{arg}(B_{\frac{1}{2},0}^{\pi\pi}) 
=  \mathrm{arg}(A_{\frac{1}{2},0}^{KK}) - \mathrm{arg}(B_{\frac{1}{2},0}^{KK}) \,. \label{eq:resonance-phase} 
\end{align}
Consequently, Eq.~(\ref{eq:amplitude-ratio}) eliminates altogether two absolute values and one phase. It 
introduces a correlation between the $D\rightarrow \pi\pi$ and $D\rightarrow KK$ systems of decays which is absent in the plain isospin parametrization.

We stress that for the implementation of the resonance models we employ 
Eqs.~(\ref{eq:amplitude-ratio}, \ref{eq:resonance-phase}) only, i.e. we do not use a specific resonance shape. 
However, we emphasize also that nevertheless we are considering simplified scenarios only, 
which are most probably a rough approximation 
of physical reality. The reason for considering the simplified scenarios is that, as long as the 
hadronic matrix elements  $M_{f_0}^{sd}$ and $M_{f_0}^{b}$ are unknown, a complete general two-resonance model 
would not lead to a reduction of complexity. The four $I=0$ amplitudes would then be merely replaced by four unknown hadronic matrix elements of the resonances and the $D^0$ meson, amounting to a fit that is equivalent to the plain isospin fit. 

Since the mass region near the charm meson mass is known from lattice studies to be very rich in gluonic content~\cite{MP99} through non-perturbative 
effects we should expect this to become a source for large breaking of SU(3)$_F$ or for that matter $U$-spin~\cite{CS84}. 
Thus this should help us understand the large breaking of SU(3)$_F$ seen in experiments since a long time ago; for example in the ratio of branching ratios for $D^0 \to K^+K^-$ versus $D^0 \to \pi^+ \pi^-$~\cite{Zyla:2020zbs}. Moreover, this characteristic of large SU(3)$_F$ breaking is also displayed in the decays of the two scalar resonances under study in this work.

\section{Numerical Results \label{sec:predictions}}

\subsection{Input Data}

The employed numerical input is summarized in Tables~\ref{tab:resonances} and \ref{tab:numericalinput}.
We calculate the direct CP asymmetries from the time-integrated ones according to Eq.~(\ref{eq:indirect}),
using the HFLAV average of indirect charm CP violation~\cite{Amhis:2019ckw, BaBar:2012bho, CDF:2014wyb, LHCb:2015xyd, Belle:2015etc, LHCb:2017ejh, Aaij:2019kcg, BaBar:2007tfw, CDF:2012ous, LHCb:2014kcb, LHCb:2016csn, LHCb:2019hro}
\begin{align}
a_{CP}^{\mathrm{ind}} &= -0.00010\pm 0.00012\,. \label{eq:indirect-charm-CPV}
\end{align}
The current world average of time-integrated CP violation in $D^0\rightarrow K_SK_S$ is given as 
$A_{CP}(D^0\rightarrow K_SK_S) = -0.019\pm 0.010$~\cite{LHCb:2021rdn, Dash:2017heu, LHCb:2015ope, CLEO:2000opx}. 
Compared to that we neglect charm and kaon indirect CP violation and 
use the approximation 
$A_{CP}(D^0\rightarrow K_SK_S) \approx a_{CP}^{\mathrm{dir}}(D^0\rightarrow K_SK_S)$. 
The branching ratio input is taken from Ref.~\cite{Zyla:2020zbs}.
We do not take into account experimental correlations between CP asymmetries or branching ratios.
For example, once the LHCb~Run~1 measurement of $A_{CP}(D^0\rightarrow K^+K^-)$ provided in Ref.~\cite{LHCb:2016nxk} is updated in the future, it is desirable to include the corresponding correlation with the LHCb Run~1+2 result for $\Delta A_{CP}$~\cite{Aaij:2019kcg}. 

\begin{table}[t]
\begin{center}
\begin{tabular}{c|c|c|c|c}
\hline \hline
Resonance                       &  $I^G(J^{PC})$ & mass $m$ [MeV]      & $\Gamma$ [MeV]     & Ref.   \\\hline 
$f_0(1710)$                     &  $0^+(0^{++})$ & $1704\pm 12$        & $123\pm 18$        & \cite{Zyla:2020zbs} \\
$f_0(1790)$                     &  $0^+(0^{++})$ & $1790^{+40}_{-30}$  & $270^{+60}_{-30}$  & \cite{Aaij:2014emv, Ablikim:2004wn} \\\hline\hline
\end{tabular}
\caption{Employed experimental data for scalar unflavored resonances close to the $D^0$ mass. 
\label{tab:resonances}}
\end{center}
\end{table}

\begin{table}[t]
\begin{center}
\begin{tabular}{c|c|c}
\hline \hline
Observable        &    Input       & Ref.  \\\hline
$\Delta a_{CP}^{\mathrm{dir}}$   & $-0.00161\pm 0.00028$   & \cite{Amhis:2019ckw, Aaij:2019kcg, BaBar:2007tfw, CDF:2012ous, LHCb:2014kcb, LHCb:2016csn, LHCb:2019hro}  \\
$a_{CP}^{\mathrm{dir}}(D^0\rightarrow \pi^+\pi^-)$ & $ 0.005\pm 0.005$  & \cite{Belle:2008ddg, CLEO:2001lgl, FOCUS:2000ejh, E791:1997txw} \\ 
$a_{CP}^{\mathrm{dir}}(D^0\rightarrow \pi^0\pi^0)$ & $-0.0002 \pm 0.006$  &  \cite{Belle:2014evd, CLEO:2000opx}   \\         
$a_{CP}^{\mathrm{dir}}(D^0\rightarrow K^+K^-)$     & $0.000 \pm 0.001$  &  \cite{LHCb:2016nxk, Belle:2008ddg, CLEO:2001lgl, FOCUS:2000ejh, E791:1997txw}   \\ 
$a_{CP}^{\mathrm{dir}}(D^+\rightarrow K^+K_S)$     &  $-0.0001 \pm 0.0007$     &  \cite{LHCb:2019dwr, BaBar:2012wep, Belle:2012ygx}  \\ 
$a_{CP}^{\mathrm{dir}}(D^0\rightarrow K_SK_S)$     &  $-0.019 \pm 0.010$    & \cite{LHCb:2021rdn, Dash:2017heu, LHCb:2015ope, CLEO:2000opx}   \\\hline\hline
\end{tabular}
\caption{Numerical input for direct CP asymmetries from experiment. 
In order to avoid double counting of data, we only use input for the single CP asymmetries
$a_{CP}^{\mathrm{dir}}(D^0\rightarrow \pi^+\pi^-)$ and $a_{CP}^{\mathrm{dir}}(D^0\rightarrow K^+K^-)$
which are not included in the HFLAV  average for $\Delta a_{CP}^{\mathrm{dir}}$, or where the corresponding 
correlation is small, namely for the LHCb Run~1 measurement of $A_{CP}(D^0\rightarrow K^+K^-)$ given in Ref.~\cite{LHCb:2016nxk}. 
Note that for this reason we do not include the result of Ref.~\cite{LHCb:2016nxk} for $A_{CP}(D^0\rightarrow \pi^+\pi^-)$.
In these cases we calculate our own corresponding averages neglecting any correlations.
\label{tab:numericalinput}}
\end{center}
\end{table}

\subsection{Fit results}

\begin{table}[t]
\begin{center}
\begin{tabular}{c|c|c|c|c}
\hline \hline
Scenario       &    
Involved resonances & 
dof &
$\Delta \chi^2$ &
Rejection/compatibility \\\hline
(1) & only $f_0(1710)$  & 3 & $\Delta \chi^2_{\mathrm{min}} = 62.5$ & $7.4\sigma$ \\
(2) & only $f_0(1790)$  & 2 & $\Delta \chi^2_{\mathrm{min}} = 3.6$ & $1.4\sigma$ \\
(3) & $f_0(1710)$ and $f_0(1790)$ & 2 & $\Delta \chi^2_{\mathrm{min}} = 3.6$ & $1.4\sigma$ \\\hline\hline
\end{tabular}
\caption{Results of the likelihood ratio tests of the simplified scenarios of the resonance model compared to the null hypothesis, which is the plain isospin fit. Due to the lack of $f_0(1790)$ branching ratio data, scenarios (2) and (3) are equivalent and give identical results. Both scenarios (2) and (3) are compatible with the data at $1.4\sigma$, whereas scenario (1) is rejected at $7.4\sigma$. 
\label{tab:scenario-tests}}
\end{center}
\end{table}

\begin{table}[t]
\begin{center}
\begin{tabular}{c|c|c}
\hline \hline
Observable        &     
Isospin &
Scalar resonance model \\\hline
$\Delta a_{CP}^{\mathrm{dir}}$  
	& $ -0.0016\pm 0.0003$
	& $ -0.0016\pm 0.0003$ 
	\\ 
$a_{CP}^{\mathrm{dir}}(D^0\rightarrow \pi^+\pi^-)$ 
	& $ 0.002 \pm 0.001$
	& $ 0.002 \pm 0.001$ 
	 \\ 
$a_{CP}^{\mathrm{dir}}(D^0\rightarrow K^+K^-)$     
	& $ 0.000\pm 0.001$ 
	& $ 0.0002^{+0.0009}_{-0.0010}$\\ 
$a_{CP}^{\mathrm{dir}}(D^0\rightarrow \pi^0\pi^0)$ 
	&  $ 0.000\pm 0.006$   
	&  $-0.005\pm 0.003$ 
	 \\ 
$a_{CP}^{\mathrm{dir}}(D^+\rightarrow K^+K_S)$     
	& $ -0.0001\pm 0.0007$  
	& $ -0.0001\pm 0.0007$   
	 \\
$a_{CP}^{\mathrm{dir}}(D^0\rightarrow K_SK_S)$     
	& $ -0.019\pm 0.010$   
	& $ -0.015^{+0.009}_{-0.010}$  
	\\\hline\hline
\end{tabular}
\caption{Theory results in the plain isospin fit as well as in the resonance model, for the simplified scenarios (2) and (3),  which give identical results. Scenario~(1) is excluded by experimental data and therefore not shown.
\label{tab:numericaloutput}}
\end{center}
\end{table}

\begin{figure}[t]
\begin{center}
\includegraphics[width=\textwidth]{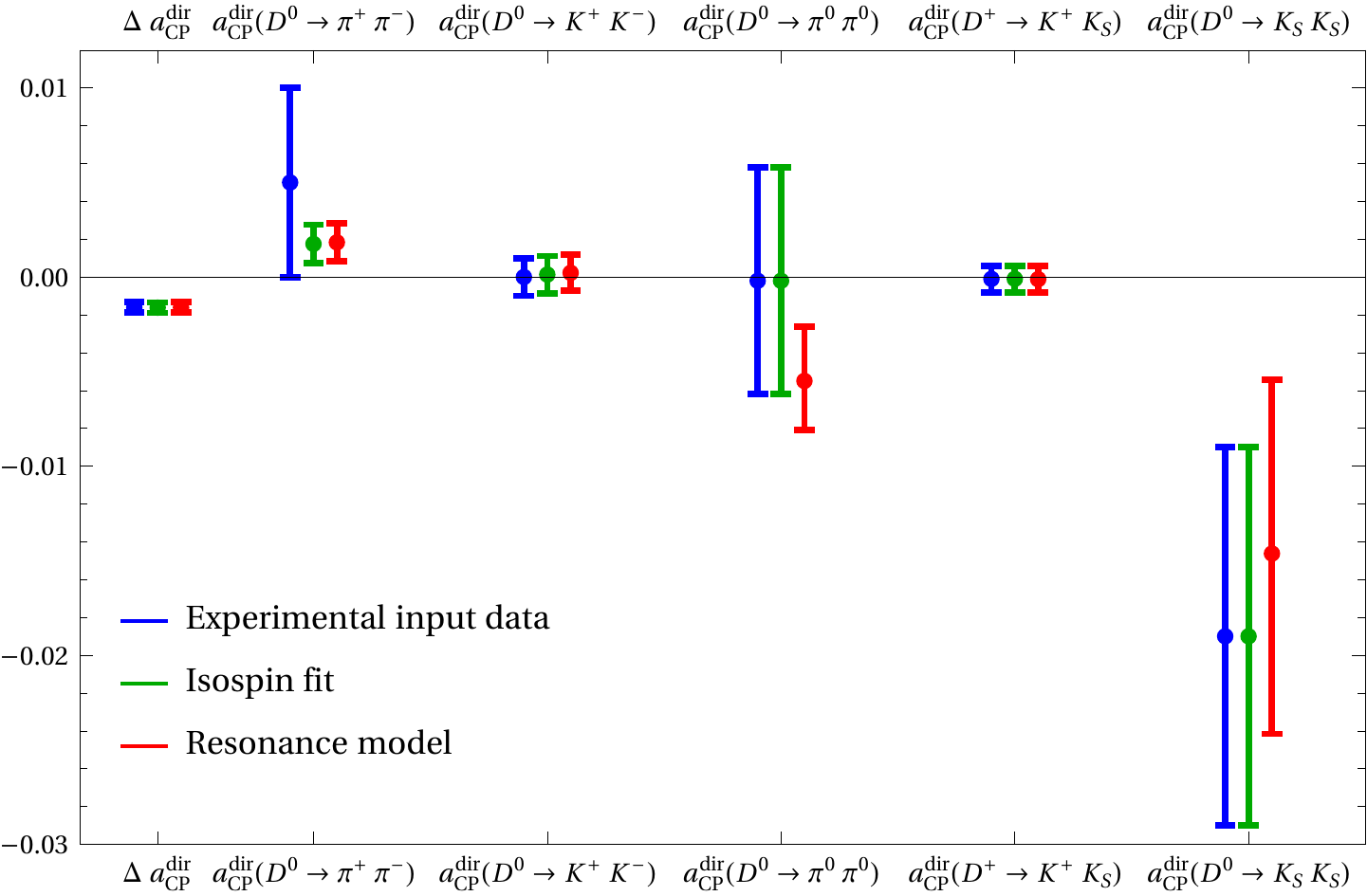}
\caption{Comparison of input data (blue), see Table~\ref{tab:numericalinput}, plain isospin fit (green) and the resonance model (red) in case of the simplified scenarios (2) and (3), which give identical results. Shown are the central values and respective $1\sigma$ error bars.
The simplified scenario (1) is excluded and therefore not shown, because in this scenario Eqs.~(\ref{eq:amplitude-ratio})--(\ref{eq:ratio-1710}) are incompatible with the charm decay data.} 
\label{fig:plot}
\end{center}
\end{figure}

For the model-independent plain isospin fit of Eqs.~(\ref{eq:isospin-first})--(\ref{eq:isospin-last}) we obtain the global minimum $\chi^2_{\mathrm{min}} = 0.4$. 
We extract in this fit scenario the charm $\Delta I=1/2$ rule, i.e.~the ratio of 
$\Delta I=1/2$ over $\Delta I=3/2$ matrix elements, as 
\begin{align}
\left| \frac{
		A_{\frac{1}{2},0}^{\pi\pi}
	}{
		A_{\frac{3}{2},2}^{\pi\pi}
	} \right| = 2.44\pm 0.04\,, \label{eq:DeltaI12-rule}
\end{align}
in agreement with extractions in Refs.~\cite{Franco:2012ck, CLEO:2005mti}. 

From the perspective of the resonance model, Eq.~(\ref{eq:DeltaI12-rule}) reflects the  
dominance of charm decays by scalar resonances close to the $D^0$ mass.
The scalar resonance models are implemented by imposing Eqs.~(\ref{eq:amplitude-ratio}, \ref{eq:resonance-phase}) on the isospin fit. Thereby, the resonance models implicitly include an assumption about the mechanism of SU(3)$_F$ breaking. 
The relative degrees of freedom of the resonance models compared to the plain isospin fit, \emph{i.e.},~the number of parameters fixed through Eqs.~(\ref{eq:amplitude-ratio}, \ref{eq:resonance-phase}), is $\nu=3$ in scenario (1), and $\nu=2$ for scenarios (2) and (3), as in these scenarios the ratio $r_{f_0}$ is left unconstrained. 

The results of the likelihood ratio tests of the three scenarios compared to the null hypothesis, which is the 
plain isospin fit, are shown in Table~\ref{tab:scenario-tests}.
The model based solely on the $f_0(1710)$ is excluded at $7.4\sigma$.
The scenarios including only the $f_0(1790)$ and both $f_0(1710)$ and $f_0(1790)$, are currently both a viable 
description of the data.
Note however, that this is mainly because in case of the $f_0(1710)$ we have a measurement of the ratio of branching ratios
Eq.~(\ref{eq:ratio-1710}), which is not the case for the $f_0(1790)$ resonance. 

In scenario (3), i.e.~when combining $f_0(1790)$ and $f_0(1710)$, we obtain
\begin{align}
\vert r_{f_0}\vert &= 0.74^{+0.06}_{-0.02}\,, \label{eq:rf0-sc3}   
\end{align}
with $r_{f_0}$ defined in Eq.~(\ref{eq:amplitude-ratio}).
Note that the limit $\vert r_{f_0}\vert = 1$ corresponds to no SU(3)$_F$-breaking between the $I=0$ amplitudes.
Therefore, Eq.~(\ref{eq:rf0-sc3}) implies SU(3)$_F$-breaking of $\sim 30\%$, in agreement with 
model-independent studies~\cite{Hiller:2012xm, Muller:2015lua}.
The further interpretation of Eq.~(\ref{eq:rf0-sc3}) depends on the details of the considered resonance model. 

In scenario (2), i.e.~the model based on the $f_0(1790)$ resonance only, we can predict the ratio of branching ratios to pions and kaons needed in order to explain the charm decay data. We obtain 
\begin{align}
\frac{\mathcal{B}(f_0(1790)\rightarrow \pi\pi)}{\mathcal{B}(f_0(1790)\rightarrow KK)} &=  0.65^{+0.12}_{-0.04}\,. \label{eq:ratio-prediction}  
\end{align}
More branching ratio data for the $f_0(1790)$ resonance could in principle easily challenge scenario (2). 
In the literature, we find only the qualitative observation, that $f_0(1790)\rightarrow  KK$ decays are relatively suppressed compared to $f_0(1790)\rightarrow \pi\pi$~\cite{Ablikim:2004wn}. If confirmed, this could possibly lead to the exclusion of scenario (2) and make more complicated scenarios necessary, such as scenario (3). 
We encourage dedicated measurements of the branching ratios of the $f_0(1790)$ resonance in order to test our model.

In Table~\ref{tab:numericaloutput} and Fig.~\ref{fig:plot} we show our theory results for the CP asymmetries 
obtained from the isospin fit and the resonance model in scenarios (2) and (3), compared to the input data.
The isospin fit basically just reproduces the data.
As in both scenarios (2) and (3) the ratio $r_{f_0}$ is unknown, these two scenarios give currently identical results. 
We obtain nontrivial theory results especially 
in case of the decays $D^0\rightarrow K_SK_S$ and $D^0\rightarrow \pi^0\pi^0$. Notably we predict  
the sign of $a_{CP}^{\mathrm{dir}}(D^0\rightarrow \pi^0\pi^0)$ to be negative. 
Furthermore, the resonance model seems to indicate that out of the final states of $D$ decays (see Table~\ref{tab:numericaloutput}) that we are considering, $a_{CP}^{\mathrm{dir}}(D^0\rightarrow K_SK_S)$ is the largest CP asymmetry, in agreement with the tendency seen in model-independent predictions~\cite{Atwood:2012ac,Nierste:2015zra}, followed by $a_{CP}^{\mathrm{dir}}(D^0\rightarrow \pi^0\pi^0)$. 

Note that due to the presence of resonances, in our model deviations from the $U$-spin symmetry are possible.
Due to the large uncertainties regarding the information about the scalar resonances, the theoretical results are currently very broad. 
Still, with more precise CP asymmetry measurements, as well as with branching ratio measurements of the scalar resonances, in the future it will become feasible to test the model directly from data, and to distinguish it from other approaches.

\section{Conclusions \label{sec:conclusions}}

Modelling the $KK$--$\pi\pi$ rescattering in charm decays with scalar resonances close to the neutral charm meson mass,
we can reasonably well account for the recently measured charm CP violation within the Standard Model. 
Thus, from the perspective of this resonance model there does not seem to be a compelling need for new physics at this time.
Our theory results for further charm CP asymmetries make it possible to test our proposed dynamical mechanism with 
future measurements. More experimental information on the scalar $f_0$ resonances will 
improve the theoretical results for charm CP violation. 
In particular, it is extremely important to measure their branching ratios to $KK$ and $\pi\pi$ modes as precisely as possible. 
Lattice studies in the charm region should also help.
We strongly encourage experimentalists to measure all CP asymmetries of singly-Cabibbo suppressed charm decays in order to test  
the different theoretical scenarios being proposed and to search for new physics.

\begin{acknowledgments}
We thank Avital Dery, Marco Gersabeck, Tim Gershon, Yuval Grossman, Alexander Lenz, Michael Morello, Ulrich Nierste, Tommaso Pajero, Aleksey Rusov and Sheldon Stone
for discussions. We also want to thank Sheldon Stone for pointing out the resonance $f_0(1790)$.
S.S. is supported by a Stephen Hawking Fellowship from UKRI under reference EP/T01623X/1 and the Lancaster-Manchester-Sheffield Consortium for Fundamental Physics, under STFC research grant ST/T001038/1.
The work of A.S.~is supported in part by the US DOE Contract No. DE-SC 0012704.
\end{acknowledgments}

\bibliography{draft.bib}

\begin{thebibliography}{112}%
\makeatletter
\providecommand \@ifxundefined [1]{%
 \@ifx{#1\undefined}
}%
\providecommand \@ifnum [1]{%
 \ifnum #1\expandafter \@firstoftwo
 \else \expandafter \@secondoftwo
 \fi
}%
\providecommand \@ifx [1]{%
 \ifx #1\expandafter \@firstoftwo
 \else \expandafter \@secondoftwo
 \fi
}%
\providecommand \natexlab [1]{#1}%
\providecommand \enquote  [1]{``#1''}%
\providecommand \bibnamefont  [1]{#1}%
\providecommand \bibfnamefont [1]{#1}%
\providecommand \citenamefont [1]{#1}%
\providecommand \href@noop [0]{\@secondoftwo}%
\providecommand \href [0]{\begingroup \@sanitize@url \@href}%
\providecommand \@href[1]{\@@startlink{#1}\@@href}%
\providecommand \@@href[1]{\endgroup#1\@@endlink}%
\providecommand \@sanitize@url [0]{\catcode `\\12\catcode `\$12\catcode
  `\&12\catcode `\#12\catcode `\^12\catcode `\_12\catcode `\%12\relax}%
\providecommand \@@startlink[1]{}%
\providecommand \@@endlink[0]{}%
\providecommand \url  [0]{\begingroup\@sanitize@url \@url }%
\providecommand \@url [1]{\endgroup\@href {#1}{\urlprefix }}%
\providecommand \urlprefix  [0]{URL }%
\providecommand \Eprint [0]{\href }%
\providecommand \doibase [0]{http://dx.doi.org/}%
\providecommand \selectlanguage [0]{\@gobble}%
\providecommand \bibinfo  [0]{\@secondoftwo}%
\providecommand \bibfield  [0]{\@secondoftwo}%
\providecommand \translation [1]{[#1]}%
\providecommand \BibitemOpen [0]{}%
\providecommand \bibitemStop [0]{}%
\providecommand \bibitemNoStop [0]{.\EOS\space}%
\providecommand \EOS [0]{\spacefactor3000\relax}%
\providecommand \BibitemShut  [1]{\csname bibitem#1\endcsname}%
\let\auto@bib@innerbib\@empty
\bibitem [{\citenamefont {Christenson}\ \emph {et~al.}(1964)\citenamefont
  {Christenson}, \citenamefont {Cronin}, \citenamefont {Fitch},\ and\
  \citenamefont {Turlay}}]{Christenson:1964fg}%
  \BibitemOpen
  \bibfield  {author} {\bibinfo {author} {\bibfnamefont {J.~H.}\ \bibnamefont
  {Christenson}}, \bibinfo {author} {\bibfnamefont {J.~W.}\ \bibnamefont
  {Cronin}}, \bibinfo {author} {\bibfnamefont {V.~L.}\ \bibnamefont {Fitch}}, \
  and\ \bibinfo {author} {\bibfnamefont {R.}~\bibnamefont {Turlay}},\ }\href
  {\doibase 10.1103/PhysRevLett.13.138} {\bibfield  {journal} {\bibinfo
  {journal} {Phys. Rev. Lett.}\ }\textbf {\bibinfo {volume} {13}},\ \bibinfo
  {pages} {138} (\bibinfo {year} {1964})}\BibitemShut {NoStop}%
\bibitem [{\citenamefont {Abbott}\ \emph {et~al.}(2020)\citenamefont {Abbott}
  \emph {et~al.}}]{Abbott:2020hxn}%
  \BibitemOpen
  \bibfield  {author} {\bibinfo {author} {\bibfnamefont {R.}~\bibnamefont
  {Abbott}} \emph {et~al.} (\bibinfo {collaboration} {RBC, UKQCD}),\ }\href
  {\doibase 10.1103/PhysRevD.102.054509} {\bibfield  {journal} {\bibinfo
  {journal} {Phys. Rev.}\ }\textbf {\bibinfo {volume} {D102}},\ \bibinfo
  {pages} {054509} (\bibinfo {year} {2020})},\ \Eprint
  {http://arxiv.org/abs/2004.09440} {arXiv:2004.09440 [hep-lat]} \BibitemShut
  {NoStop}%
\bibitem [{\citenamefont {Aebischer}\ \emph {et~al.}(2020)\citenamefont
  {Aebischer}, \citenamefont {Bobeth},\ and\ \citenamefont
  {Buras}}]{Aebischer:2020jto}%
  \BibitemOpen
  \bibfield  {author} {\bibinfo {author} {\bibfnamefont {J.}~\bibnamefont
  {Aebischer}}, \bibinfo {author} {\bibfnamefont {C.}~\bibnamefont {Bobeth}}, \
  and\ \bibinfo {author} {\bibfnamefont {A.~J.}\ \bibnamefont {Buras}},\ }\href
  {\doibase 10.1140/epjc/s10052-020-8267-1} {\bibfield  {journal} {\bibinfo
  {journal} {Eur. Phys. J.}\ }\textbf {\bibinfo {volume} {C80}},\ \bibinfo
  {pages} {705} (\bibinfo {year} {2020})},\ \Eprint
  {http://arxiv.org/abs/2005.05978} {arXiv:2005.05978 [hep-ph]} \BibitemShut
  {NoStop}%
\bibitem [{\citenamefont {Cirigliano}\ \emph {et~al.}(2020)\citenamefont
  {Cirigliano}, \citenamefont {Gisbert}, \citenamefont {Pich},\ and\
  \citenamefont {Rodríguez-Sánchez}}]{Cirigliano:2019ani}%
  \BibitemOpen
  \bibfield  {author} {\bibinfo {author} {\bibfnamefont {V.}~\bibnamefont
  {Cirigliano}}, \bibinfo {author} {\bibfnamefont {H.}~\bibnamefont {Gisbert}},
  \bibinfo {author} {\bibfnamefont {A.}~\bibnamefont {Pich}}, \ and\ \bibinfo
  {author} {\bibfnamefont {A.}~\bibnamefont {Rodríguez-Sánchez}},\ }\bibfield
   {booktitle} {\emph {\bibinfo {booktitle} {{Proceedings, International
  Conference on Kaon Physics 2019 (KAON2019): Perugia, Italy, September 10-13,
  2019}}},\ }\href {\doibase 10.1088/1742-6596/1526/1/012011} {\bibfield
  {journal} {\bibinfo  {journal} {J. Phys. Conf. Ser.}\ }\textbf {\bibinfo
  {volume} {1526}},\ \bibinfo {pages} {012011} (\bibinfo {year} {2020})},\
  \Eprint {http://arxiv.org/abs/1912.04736} {arXiv:1912.04736 [hep-ph]}
  \BibitemShut {NoStop}%
\bibitem [{\citenamefont {Zyla}\ \emph {et~al.}(2020)\citenamefont {Zyla} \emph
  {et~al.}}]{Zyla:2020zbs}%
  \BibitemOpen
  \bibfield  {author} {\bibinfo {author} {\bibfnamefont {P.~A.}\ \bibnamefont
  {Zyla}} \emph {et~al.} (\bibinfo {collaboration} {Particle Data Group}),\
  }\href {\doibase 10.1093/ptep/ptaa104} {\bibfield  {journal} {\bibinfo
  {journal} {PTEP}\ }\textbf {\bibinfo {volume} {2020}},\ \bibinfo {pages}
  {083C01} (\bibinfo {year} {2020})}\BibitemShut {NoStop}%
\bibitem [{\citenamefont {Aaij}\ \emph
  {et~al.}(2021{\natexlab{a}})\citenamefont {Aaij} \emph
  {et~al.}}]{Aaij:2020wnj}%
  \BibitemOpen
  \bibfield  {author} {\bibinfo {author} {\bibfnamefont {R.}~\bibnamefont
  {Aaij}} \emph {et~al.} (\bibinfo {collaboration} {LHCb}),\ }\href {\doibase
  10.1103/PhysRevLett.126.091802} {\bibfield  {journal} {\bibinfo  {journal}
  {Phys. Rev. Lett.}\ }\textbf {\bibinfo {volume} {126}},\ \bibinfo {pages}
  {091802} (\bibinfo {year} {2021}{\natexlab{a}})},\ \Eprint
  {http://arxiv.org/abs/2012.12789} {arXiv:2012.12789 [hep-ex]} \BibitemShut
  {NoStop}%
\bibitem [{\citenamefont {Amhis}\ \emph {et~al.}(2021)\citenamefont {Amhis}
  \emph {et~al.}}]{Amhis:2019ckw}%
  \BibitemOpen
  \bibfield  {author} {\bibinfo {author} {\bibfnamefont {Y.~S.}\ \bibnamefont
  {Amhis}} \emph {et~al.} (\bibinfo {collaboration} {HFLAV}),\ }\href {\doibase
  10.1140/epjc/s10052-020-8156-7} {\bibfield  {journal} {\bibinfo  {journal}
  {Eur. Phys. J. C}\ }\textbf {\bibinfo {volume} {81}},\ \bibinfo {pages} {226}
  (\bibinfo {year} {2021})},\ \Eprint {http://arxiv.org/abs/1909.12524}
  {arXiv:1909.12524 [hep-ex]} \BibitemShut {NoStop}%
\bibitem [{\citenamefont {Aaij}\ \emph
  {et~al.}(2019{\natexlab{a}})\citenamefont {Aaij} \emph
  {et~al.}}]{Aaij:2019kcg}%
  \BibitemOpen
  \bibfield  {author} {\bibinfo {author} {\bibfnamefont {R.}~\bibnamefont
  {Aaij}} \emph {et~al.} (\bibinfo {collaboration} {LHCb}),\ }\href {\doibase
  10.1103/PhysRevLett.122.211803} {\bibfield  {journal} {\bibinfo  {journal}
  {Phys. Rev. Lett.}\ }\textbf {\bibinfo {volume} {122}},\ \bibinfo {pages}
  {211803} (\bibinfo {year} {2019}{\natexlab{a}})},\ \Eprint
  {http://arxiv.org/abs/1903.08726} {arXiv:1903.08726 [hep-ex]} \BibitemShut
  {NoStop}%
\bibitem [{\citenamefont {Aubert}\ \emph {et~al.}(2008)\citenamefont {Aubert}
  \emph {et~al.}}]{BaBar:2007tfw}%
  \BibitemOpen
  \bibfield  {author} {\bibinfo {author} {\bibfnamefont {B.}~\bibnamefont
  {Aubert}} \emph {et~al.} (\bibinfo {collaboration} {BaBar}),\ }\href
  {\doibase 10.1103/PhysRevLett.100.061803} {\bibfield  {journal} {\bibinfo
  {journal} {Phys. Rev. Lett.}\ }\textbf {\bibinfo {volume} {100}},\ \bibinfo
  {pages} {061803} (\bibinfo {year} {2008})},\ \Eprint
  {http://arxiv.org/abs/0709.2715} {arXiv:0709.2715 [hep-ex]} \BibitemShut
  {NoStop}%
\bibitem [{\citenamefont {Aaltonen}\ \emph {et~al.}(2012)\citenamefont
  {Aaltonen} \emph {et~al.}}]{CDF:2012ous}%
  \BibitemOpen
  \bibfield  {author} {\bibinfo {author} {\bibfnamefont {T.}~\bibnamefont
  {Aaltonen}} \emph {et~al.} (\bibinfo {collaboration} {CDF}),\ }\href
  {\doibase 10.1103/PhysRevLett.109.111801} {\bibfield  {journal} {\bibinfo
  {journal} {Phys. Rev. Lett.}\ }\textbf {\bibinfo {volume} {109}},\ \bibinfo
  {pages} {111801} (\bibinfo {year} {2012})},\ \Eprint
  {http://arxiv.org/abs/1207.2158} {arXiv:1207.2158 [hep-ex]} \BibitemShut
  {NoStop}%
\bibitem [{\citenamefont {Aaij}\ \emph
  {et~al.}(2014{\natexlab{a}})\citenamefont {Aaij} \emph
  {et~al.}}]{LHCb:2014kcb}%
  \BibitemOpen
  \bibfield  {author} {\bibinfo {author} {\bibfnamefont {R.}~\bibnamefont
  {Aaij}} \emph {et~al.} (\bibinfo {collaboration} {LHCb}),\ }\href {\doibase
  10.1007/JHEP07(2014)041} {\bibfield  {journal} {\bibinfo  {journal} {JHEP}\
  }\textbf {\bibinfo {volume} {07}},\ \bibinfo {pages} {041} (\bibinfo {year}
  {2014}{\natexlab{a}})},\ \Eprint {http://arxiv.org/abs/1405.2797}
  {arXiv:1405.2797 [hep-ex]} \BibitemShut {NoStop}%
\bibitem [{\citenamefont {Aaij}\ \emph {et~al.}(2016)\citenamefont {Aaij} \emph
  {et~al.}}]{LHCb:2016csn}%
  \BibitemOpen
  \bibfield  {author} {\bibinfo {author} {\bibfnamefont {R.}~\bibnamefont
  {Aaij}} \emph {et~al.} (\bibinfo {collaboration} {LHCb}),\ }\href {\doibase
  10.1103/PhysRevLett.116.191601} {\bibfield  {journal} {\bibinfo  {journal}
  {Phys. Rev. Lett.}\ }\textbf {\bibinfo {volume} {116}},\ \bibinfo {pages}
  {191601} (\bibinfo {year} {2016})},\ \Eprint
  {http://arxiv.org/abs/1602.03160} {arXiv:1602.03160 [hep-ex]} \BibitemShut
  {NoStop}%
\bibitem [{\citenamefont {Aaij}\ \emph
  {et~al.}(2019{\natexlab{b}})\citenamefont {Aaij} \emph
  {et~al.}}]{LHCb:2019hro}%
  \BibitemOpen
  \bibfield  {author} {\bibinfo {author} {\bibfnamefont {R.}~\bibnamefont
  {Aaij}} \emph {et~al.} (\bibinfo {collaboration} {LHCb}),\ }\href {\doibase
  10.1103/PhysRevLett.122.211803} {\bibfield  {journal} {\bibinfo  {journal}
  {Phys. Rev. Lett.}\ }\textbf {\bibinfo {volume} {122}},\ \bibinfo {pages}
  {211803} (\bibinfo {year} {2019}{\natexlab{b}})},\ \Eprint
  {http://arxiv.org/abs/1903.08726} {arXiv:1903.08726 [hep-ex]} \BibitemShut
  {NoStop}%
\bibitem [{\citenamefont {Grossman}\ \emph {et~al.}(2007)\citenamefont
  {Grossman}, \citenamefont {Kagan},\ and\ \citenamefont
  {Nir}}]{Grossman:2006jg}%
  \BibitemOpen
  \bibfield  {author} {\bibinfo {author} {\bibfnamefont {Y.}~\bibnamefont
  {Grossman}}, \bibinfo {author} {\bibfnamefont {A.~L.}\ \bibnamefont {Kagan}},
  \ and\ \bibinfo {author} {\bibfnamefont {Y.}~\bibnamefont {Nir}},\ }\href
  {\doibase 10.1103/PhysRevD.75.036008} {\bibfield  {journal} {\bibinfo
  {journal} {Phys. Rev. D}\ }\textbf {\bibinfo {volume} {75}},\ \bibinfo
  {pages} {036008} (\bibinfo {year} {2007})},\ \Eprint
  {http://arxiv.org/abs/hep-ph/0609178} {arXiv:hep-ph/0609178} \BibitemShut
  {NoStop}%
\bibitem [{\citenamefont {Aaij}\ \emph
  {et~al.}(2021{\natexlab{b}})\citenamefont {Aaij} \emph
  {et~al.}}]{LHCb:2021rdn}%
  \BibitemOpen
  \bibfield  {author} {\bibinfo {author} {\bibfnamefont {R.}~\bibnamefont
  {Aaij}} \emph {et~al.} (\bibinfo {collaboration} {LHCb}),\ }\href@noop {} {\
  (\bibinfo {year} {2021}{\natexlab{b}})},\ \Eprint
  {http://arxiv.org/abs/2105.01565} {arXiv:2105.01565 [hep-ex]} \BibitemShut
  {NoStop}%
\bibitem [{\citenamefont {Dash}\ \emph {et~al.}(2017)\citenamefont {Dash} \emph
  {et~al.}}]{Dash:2017heu}%
  \BibitemOpen
  \bibfield  {author} {\bibinfo {author} {\bibfnamefont {N.}~\bibnamefont
  {Dash}} \emph {et~al.},\ }\href {\doibase 10.1103/PhysRevLett.119.171801}
  {\bibfield  {journal} {\bibinfo  {journal} {Phys. Rev. Lett.}\ }\textbf
  {\bibinfo {volume} {119}},\ \bibinfo {pages} {171801} (\bibinfo {year}
  {2017})},\ \Eprint {http://arxiv.org/abs/1705.05966} {arXiv:1705.05966
  [hep-ex]} \BibitemShut {NoStop}%
\bibitem [{\citenamefont {Aaij}\ \emph
  {et~al.}(2015{\natexlab{a}})\citenamefont {Aaij} \emph
  {et~al.}}]{LHCb:2015ope}%
  \BibitemOpen
  \bibfield  {author} {\bibinfo {author} {\bibfnamefont {R.}~\bibnamefont
  {Aaij}} \emph {et~al.} (\bibinfo {collaboration} {LHCb}),\ }\href {\doibase
  10.1007/JHEP10(2015)055} {\bibfield  {journal} {\bibinfo  {journal} {JHEP}\
  }\textbf {\bibinfo {volume} {10}},\ \bibinfo {pages} {055} (\bibinfo {year}
  {2015}{\natexlab{a}})},\ \Eprint {http://arxiv.org/abs/1508.06087}
  {arXiv:1508.06087 [hep-ex]} \BibitemShut {NoStop}%
\bibitem [{\citenamefont {Bonvicini}\ \emph {et~al.}(2001)\citenamefont
  {Bonvicini} \emph {et~al.}}]{CLEO:2000opx}%
  \BibitemOpen
  \bibfield  {author} {\bibinfo {author} {\bibfnamefont {G.}~\bibnamefont
  {Bonvicini}} \emph {et~al.} (\bibinfo {collaboration} {CLEO}),\ }\href
  {\doibase 10.1103/PhysRevD.63.071101} {\bibfield  {journal} {\bibinfo
  {journal} {Phys. Rev. D}\ }\textbf {\bibinfo {volume} {63}},\ \bibinfo
  {pages} {071101} (\bibinfo {year} {2001})},\ \Eprint
  {http://arxiv.org/abs/hep-ex/0012054} {arXiv:hep-ex/0012054} \BibitemShut
  {NoStop}%
\bibitem [{\citenamefont {Nierste}\ and\ \citenamefont
  {Schacht}(2015)}]{Nierste:2015zra}%
  \BibitemOpen
  \bibfield  {author} {\bibinfo {author} {\bibfnamefont {U.}~\bibnamefont
  {Nierste}}\ and\ \bibinfo {author} {\bibfnamefont {S.}~\bibnamefont
  {Schacht}},\ }\href {\doibase 10.1103/PhysRevD.92.054036} {\bibfield
  {journal} {\bibinfo  {journal} {Phys. Rev.}\ }\textbf {\bibinfo {volume}
  {D92}},\ \bibinfo {pages} {054036} (\bibinfo {year} {2015})},\ \Eprint
  {http://arxiv.org/abs/1508.00074} {arXiv:1508.00074 [hep-ph]} \BibitemShut
  {NoStop}%
\bibitem [{\citenamefont {Chala}\ \emph {et~al.}(2019)\citenamefont {Chala},
  \citenamefont {Lenz}, \citenamefont {Rusov},\ and\ \citenamefont
  {Scholtz}}]{Chala:2019fdb}%
  \BibitemOpen
  \bibfield  {author} {\bibinfo {author} {\bibfnamefont {M.}~\bibnamefont
  {Chala}}, \bibinfo {author} {\bibfnamefont {A.}~\bibnamefont {Lenz}},
  \bibinfo {author} {\bibfnamefont {A.~V.}\ \bibnamefont {Rusov}}, \ and\
  \bibinfo {author} {\bibfnamefont {J.}~\bibnamefont {Scholtz}},\ }\href
  {\doibase 10.1007/JHEP07(2019)161} {\bibfield  {journal} {\bibinfo  {journal}
  {JHEP}\ }\textbf {\bibinfo {volume} {07}},\ \bibinfo {pages} {161} (\bibinfo
  {year} {2019})},\ \Eprint {http://arxiv.org/abs/1903.10490} {arXiv:1903.10490
  [hep-ph]} \BibitemShut {NoStop}%
\bibitem [{\citenamefont {Khodjamirian}\ and\ \citenamefont
  {Petrov}(2017)}]{Khodjamirian:2017zdu}%
  \BibitemOpen
  \bibfield  {author} {\bibinfo {author} {\bibfnamefont {A.}~\bibnamefont
  {Khodjamirian}}\ and\ \bibinfo {author} {\bibfnamefont {A.~A.}\ \bibnamefont
  {Petrov}},\ }\href {\doibase 10.1016/j.physletb.2017.09.070} {\bibfield
  {journal} {\bibinfo  {journal} {Phys. Lett.}\ }\textbf {\bibinfo {volume}
  {B774}},\ \bibinfo {pages} {235} (\bibinfo {year} {2017})},\ \Eprint
  {http://arxiv.org/abs/1706.07780} {arXiv:1706.07780 [hep-ph]} \BibitemShut
  {NoStop}%
\bibitem [{\citenamefont {Grossman}\ and\ \citenamefont
  {Schacht}(2019{\natexlab{a}})}]{Grossman:2019xcj}%
  \BibitemOpen
  \bibfield  {author} {\bibinfo {author} {\bibfnamefont {Y.}~\bibnamefont
  {Grossman}}\ and\ \bibinfo {author} {\bibfnamefont {S.}~\bibnamefont
  {Schacht}},\ }\href {\doibase 10.1007/JHEP07(2019)020} {\bibfield  {journal}
  {\bibinfo  {journal} {JHEP}\ }\textbf {\bibinfo {volume} {07}},\ \bibinfo
  {pages} {020} (\bibinfo {year} {2019}{\natexlab{a}})},\ \Eprint
  {http://arxiv.org/abs/1903.10952} {arXiv:1903.10952 [hep-ph]} \BibitemShut
  {NoStop}%
\bibitem [{\citenamefont {Li}\ \emph {et~al.}(2019)\citenamefont {Li},
  \citenamefont {Lü},\ and\ \citenamefont {Yu}}]{Li:2019hho}%
  \BibitemOpen
  \bibfield  {author} {\bibinfo {author} {\bibfnamefont {H.-N.}\ \bibnamefont
  {Li}}, \bibinfo {author} {\bibfnamefont {C.-D.}\ \bibnamefont {Lü}}, \ and\
  \bibinfo {author} {\bibfnamefont {F.-S.}\ \bibnamefont {Yu}},\ }\href@noop {}
  {\  (\bibinfo {year} {2019})},\ \Eprint {http://arxiv.org/abs/1903.10638}
  {arXiv:1903.10638 [hep-ph]} \BibitemShut {NoStop}%
\bibitem [{\citenamefont {Soni}(2019)}]{Soni:2019xko}%
  \BibitemOpen
  \bibfield  {author} {\bibinfo {author} {\bibfnamefont {A.}~\bibnamefont
  {Soni}},\ }\href@noop {} {\  (\bibinfo {year} {2019})},\ \Eprint
  {http://arxiv.org/abs/1905.00907} {arXiv:1905.00907 [hep-ph]} \BibitemShut
  {NoStop}%
\bibitem [{\citenamefont {Cheng}\ and\ \citenamefont
  {Chiang}(2019)}]{Cheng:2019ggx}%
  \BibitemOpen
  \bibfield  {author} {\bibinfo {author} {\bibfnamefont {H.-Y.}\ \bibnamefont
  {Cheng}}\ and\ \bibinfo {author} {\bibfnamefont {C.-W.}\ \bibnamefont
  {Chiang}},\ }\href {\doibase 10.1103/PhysRevD.100.093002} {\bibfield
  {journal} {\bibinfo  {journal} {Phys. Rev. D}\ }\textbf {\bibinfo {volume}
  {100}},\ \bibinfo {pages} {093002} (\bibinfo {year} {2019})},\ \Eprint
  {http://arxiv.org/abs/1909.03063} {arXiv:1909.03063 [hep-ph]} \BibitemShut
  {NoStop}%
\bibitem [{\citenamefont {Buras}\ \emph {et~al.}(2021)\citenamefont {Buras},
  \citenamefont {Colangelo}, \citenamefont {De~Fazio},\ and\ \citenamefont
  {Loparco}}]{Buras:2021rdg}%
  \BibitemOpen
  \bibfield  {author} {\bibinfo {author} {\bibfnamefont {A.~J.}\ \bibnamefont
  {Buras}}, \bibinfo {author} {\bibfnamefont {P.}~\bibnamefont {Colangelo}},
  \bibinfo {author} {\bibfnamefont {F.}~\bibnamefont {De~Fazio}}, \ and\
  \bibinfo {author} {\bibfnamefont {F.}~\bibnamefont {Loparco}},\ }\href@noop
  {} {\  (\bibinfo {year} {2021})},\ \Eprint {http://arxiv.org/abs/2107.10866}
  {arXiv:2107.10866 [hep-ph]} \BibitemShut {NoStop}%
\bibitem [{\citenamefont {Acaroglu}\ and\ \citenamefont
  {Blanke}(2021)}]{Acaroglu:2021qae}%
  \BibitemOpen
  \bibfield  {author} {\bibinfo {author} {\bibfnamefont {H.}~\bibnamefont
  {Acaroglu}}\ and\ \bibinfo {author} {\bibfnamefont {M.}~\bibnamefont
  {Blanke}},\ }\href@noop {} {\  (\bibinfo {year} {2021})},\ \Eprint
  {http://arxiv.org/abs/2109.10357} {arXiv:2109.10357 [hep-ph]} \BibitemShut
  {NoStop}%
\bibitem [{\citenamefont {Einhorn}\ and\ \citenamefont
  {Quigg}(1975)}]{Einhorn:1975fw}%
  \BibitemOpen
  \bibfield  {author} {\bibinfo {author} {\bibfnamefont {M.~B.}\ \bibnamefont
  {Einhorn}}\ and\ \bibinfo {author} {\bibfnamefont {C.}~\bibnamefont
  {Quigg}},\ }\href {\doibase 10.1103/PhysRevD.12.2015} {\bibfield  {journal}
  {\bibinfo  {journal} {Phys. Rev.}\ }\textbf {\bibinfo {volume} {D12}},\
  \bibinfo {pages} {2015} (\bibinfo {year} {1975})}\BibitemShut {NoStop}%
\bibitem [{\citenamefont {Abbott}\ \emph {et~al.}(1980)\citenamefont {Abbott},
  \citenamefont {Sikivie},\ and\ \citenamefont {Wise}}]{Abbott:1979fw}%
  \BibitemOpen
  \bibfield  {author} {\bibinfo {author} {\bibfnamefont {L.~F.}\ \bibnamefont
  {Abbott}}, \bibinfo {author} {\bibfnamefont {P.}~\bibnamefont {Sikivie}}, \
  and\ \bibinfo {author} {\bibfnamefont {M.~B.}\ \bibnamefont {Wise}},\ }\href
  {\doibase 10.1103/PhysRevD.21.768} {\bibfield  {journal} {\bibinfo  {journal}
  {Phys. Rev.}\ }\textbf {\bibinfo {volume} {D21}},\ \bibinfo {pages} {768}
  (\bibinfo {year} {1980})}\BibitemShut {NoStop}%
\bibitem [{\citenamefont {Golden}\ and\ \citenamefont
  {Grinstein}(1989)}]{Golden:1989qx}%
  \BibitemOpen
  \bibfield  {author} {\bibinfo {author} {\bibfnamefont {M.}~\bibnamefont
  {Golden}}\ and\ \bibinfo {author} {\bibfnamefont {B.}~\bibnamefont
  {Grinstein}},\ }\href {\doibase 10.1016/0370-2693(89)90353-5} {\bibfield
  {journal} {\bibinfo  {journal} {Phys. Lett.}\ }\textbf {\bibinfo {volume}
  {B222}},\ \bibinfo {pages} {501} (\bibinfo {year} {1989})}\BibitemShut
  {NoStop}%
\bibitem [{\citenamefont {Brod}\ \emph
  {et~al.}(2012{\natexlab{a}})\citenamefont {Brod}, \citenamefont {Grossman},
  \citenamefont {Kagan},\ and\ \citenamefont {Zupan}}]{Brod:2012ud}%
  \BibitemOpen
  \bibfield  {author} {\bibinfo {author} {\bibfnamefont {J.}~\bibnamefont
  {Brod}}, \bibinfo {author} {\bibfnamefont {Y.}~\bibnamefont {Grossman}},
  \bibinfo {author} {\bibfnamefont {A.~L.}\ \bibnamefont {Kagan}}, \ and\
  \bibinfo {author} {\bibfnamefont {J.}~\bibnamefont {Zupan}},\ }\href
  {\doibase 10.1007/JHEP10(2012)161} {\bibfield  {journal} {\bibinfo  {journal}
  {JHEP}\ }\textbf {\bibinfo {volume} {10}},\ \bibinfo {pages} {161} (\bibinfo
  {year} {2012}{\natexlab{a}})},\ \Eprint {http://arxiv.org/abs/1203.6659}
  {arXiv:1203.6659 [hep-ph]} \BibitemShut {NoStop}%
\bibitem [{\citenamefont {Bhattacharya}\ \emph {et~al.}(2012)\citenamefont
  {Bhattacharya}, \citenamefont {Gronau},\ and\ \citenamefont
  {Rosner}}]{Bhattacharya:2012ah}%
  \BibitemOpen
  \bibfield  {author} {\bibinfo {author} {\bibfnamefont {B.}~\bibnamefont
  {Bhattacharya}}, \bibinfo {author} {\bibfnamefont {M.}~\bibnamefont
  {Gronau}}, \ and\ \bibinfo {author} {\bibfnamefont {J.~L.}\ \bibnamefont
  {Rosner}},\ }\href {\doibase 10.1103/PhysRevD.85.079901,
  10.1103/PhysRevD.85.054014} {\bibfield  {journal} {\bibinfo  {journal} {Phys.
  Rev.}\ }\textbf {\bibinfo {volume} {D85}},\ \bibinfo {pages} {054014}
  (\bibinfo {year} {2012})},\ \bibinfo {note} {[Phys.
  Rev.D85,no.7,079901(2012)]},\ \Eprint {http://arxiv.org/abs/1201.2351}
  {arXiv:1201.2351 [hep-ph]} \BibitemShut {NoStop}%
\bibitem [{\citenamefont {Franco}\ \emph {et~al.}(2012)\citenamefont {Franco},
  \citenamefont {Mishima},\ and\ \citenamefont {Silvestrini}}]{Franco:2012ck}%
  \BibitemOpen
  \bibfield  {author} {\bibinfo {author} {\bibfnamefont {E.}~\bibnamefont
  {Franco}}, \bibinfo {author} {\bibfnamefont {S.}~\bibnamefont {Mishima}}, \
  and\ \bibinfo {author} {\bibfnamefont {L.}~\bibnamefont {Silvestrini}},\
  }\href {\doibase 10.1007/JHEP05(2012)140} {\bibfield  {journal} {\bibinfo
  {journal} {JHEP}\ }\textbf {\bibinfo {volume} {05}},\ \bibinfo {pages} {140}
  (\bibinfo {year} {2012})},\ \Eprint {http://arxiv.org/abs/1203.3131}
  {arXiv:1203.3131 [hep-ph]} \BibitemShut {NoStop}%
\bibitem [{\citenamefont {Hiller}\ \emph {et~al.}(2013)\citenamefont {Hiller},
  \citenamefont {Jung},\ and\ \citenamefont {Schacht}}]{Hiller:2012xm}%
  \BibitemOpen
  \bibfield  {author} {\bibinfo {author} {\bibfnamefont {G.}~\bibnamefont
  {Hiller}}, \bibinfo {author} {\bibfnamefont {M.}~\bibnamefont {Jung}}, \ and\
  \bibinfo {author} {\bibfnamefont {S.}~\bibnamefont {Schacht}},\ }\href
  {\doibase 10.1103/PhysRevD.87.014024} {\bibfield  {journal} {\bibinfo
  {journal} {Phys. Rev.}\ }\textbf {\bibinfo {volume} {D87}},\ \bibinfo {pages}
  {014024} (\bibinfo {year} {2013})},\ \Eprint {http://arxiv.org/abs/1211.3734}
  {arXiv:1211.3734 [hep-ph]} \BibitemShut {NoStop}%
\bibitem [{\citenamefont {Nierste}\ and\ \citenamefont
  {Schacht}(2017)}]{Nierste:2017cua}%
  \BibitemOpen
  \bibfield  {author} {\bibinfo {author} {\bibfnamefont {U.}~\bibnamefont
  {Nierste}}\ and\ \bibinfo {author} {\bibfnamefont {S.}~\bibnamefont
  {Schacht}},\ }\href {\doibase 10.1103/PhysRevLett.119.251801} {\bibfield
  {journal} {\bibinfo  {journal} {Phys. Rev. Lett.}\ }\textbf {\bibinfo
  {volume} {119}},\ \bibinfo {pages} {251801} (\bibinfo {year} {2017})},\
  \Eprint {http://arxiv.org/abs/1708.03572} {arXiv:1708.03572 [hep-ph]}
  \BibitemShut {NoStop}%
\bibitem [{\citenamefont {{M\"uller}}\ \emph
  {et~al.}(2015{\natexlab{a}})\citenamefont {{M\"uller}}, \citenamefont
  {Nierste},\ and\ \citenamefont {Schacht}}]{Muller:2015rna}%
  \BibitemOpen
  \bibfield  {author} {\bibinfo {author} {\bibfnamefont {S.}~\bibnamefont
  {{M\"uller}}}, \bibinfo {author} {\bibfnamefont {U.}~\bibnamefont {Nierste}},
  \ and\ \bibinfo {author} {\bibfnamefont {S.}~\bibnamefont {Schacht}},\ }\href
  {\doibase 10.1103/PhysRevLett.115.251802} {\bibfield  {journal} {\bibinfo
  {journal} {Phys. Rev. Lett.}\ }\textbf {\bibinfo {volume} {115}},\ \bibinfo
  {pages} {251802} (\bibinfo {year} {2015}{\natexlab{a}})},\ \Eprint
  {http://arxiv.org/abs/1506.04121} {arXiv:1506.04121 [hep-ph]} \BibitemShut
  {NoStop}%
\bibitem [{\citenamefont {Grossman}\ and\ \citenamefont
  {Schacht}(2019{\natexlab{b}})}]{Grossman:2018ptn}%
  \BibitemOpen
  \bibfield  {author} {\bibinfo {author} {\bibfnamefont {Y.}~\bibnamefont
  {Grossman}}\ and\ \bibinfo {author} {\bibfnamefont {S.}~\bibnamefont
  {Schacht}},\ }\href {\doibase 10.1103/PhysRevD.99.033005} {\bibfield
  {journal} {\bibinfo  {journal} {Phys. Rev.}\ }\textbf {\bibinfo {volume}
  {D99}},\ \bibinfo {pages} {033005} (\bibinfo {year} {2019}{\natexlab{b}})},\
  \Eprint {http://arxiv.org/abs/1811.11188} {arXiv:1811.11188 [hep-ph]}
  \BibitemShut {NoStop}%
\bibitem [{\citenamefont {Buccella}\ \emph {et~al.}(1995)\citenamefont
  {Buccella}, \citenamefont {Lusignoli}, \citenamefont {Miele}, \citenamefont
  {Pugliese},\ and\ \citenamefont {Santorelli}}]{Buccella:1994nf}%
  \BibitemOpen
  \bibfield  {author} {\bibinfo {author} {\bibfnamefont {F.}~\bibnamefont
  {Buccella}}, \bibinfo {author} {\bibfnamefont {M.}~\bibnamefont {Lusignoli}},
  \bibinfo {author} {\bibfnamefont {G.}~\bibnamefont {Miele}}, \bibinfo
  {author} {\bibfnamefont {A.}~\bibnamefont {Pugliese}}, \ and\ \bibinfo
  {author} {\bibfnamefont {P.}~\bibnamefont {Santorelli}},\ }\href {\doibase
  10.1103/PhysRevD.51.3478} {\bibfield  {journal} {\bibinfo  {journal} {Phys.
  Rev.}\ }\textbf {\bibinfo {volume} {D51}},\ \bibinfo {pages} {3478} (\bibinfo
  {year} {1995})},\ \Eprint {http://arxiv.org/abs/hep-ph/9411286}
  {arXiv:hep-ph/9411286 [hep-ph]} \BibitemShut {NoStop}%
\bibitem [{\citenamefont {Artuso}\ \emph {et~al.}(2008)\citenamefont {Artuso},
  \citenamefont {Meadows},\ and\ \citenamefont {Petrov}}]{Artuso:2008vf}%
  \BibitemOpen
  \bibfield  {author} {\bibinfo {author} {\bibfnamefont {M.}~\bibnamefont
  {Artuso}}, \bibinfo {author} {\bibfnamefont {B.}~\bibnamefont {Meadows}}, \
  and\ \bibinfo {author} {\bibfnamefont {A.~A.}\ \bibnamefont {Petrov}},\
  }\href {\doibase 10.1146/annurev.nucl.58.110707.171131} {\bibfield  {journal}
  {\bibinfo  {journal} {Ann. Rev. Nucl. Part. Sci.}\ }\textbf {\bibinfo
  {volume} {58}},\ \bibinfo {pages} {249} (\bibinfo {year} {2008})},\ \Eprint
  {http://arxiv.org/abs/0802.2934} {arXiv:0802.2934 [hep-ph]} \BibitemShut
  {NoStop}%
\bibitem [{\citenamefont {Buccella}\ \emph {et~al.}(2013)\citenamefont
  {Buccella}, \citenamefont {Lusignoli}, \citenamefont {Pugliese},\ and\
  \citenamefont {Santorelli}}]{Buccella:2013tya}%
  \BibitemOpen
  \bibfield  {author} {\bibinfo {author} {\bibfnamefont {F.}~\bibnamefont
  {Buccella}}, \bibinfo {author} {\bibfnamefont {M.}~\bibnamefont {Lusignoli}},
  \bibinfo {author} {\bibfnamefont {A.}~\bibnamefont {Pugliese}}, \ and\
  \bibinfo {author} {\bibfnamefont {P.}~\bibnamefont {Santorelli}},\ }\href
  {\doibase 10.1103/PhysRevD.88.074011} {\bibfield  {journal} {\bibinfo
  {journal} {Phys. Rev.}\ }\textbf {\bibinfo {volume} {D88}},\ \bibinfo {pages}
  {074011} (\bibinfo {year} {2013})},\ \Eprint {http://arxiv.org/abs/1305.7343}
  {arXiv:1305.7343 [hep-ph]} \BibitemShut {NoStop}%
\bibitem [{\citenamefont {Cheng}\ and\ \citenamefont
  {Chiang}(2012)}]{Cheng:2012wr}%
  \BibitemOpen
  \bibfield  {author} {\bibinfo {author} {\bibfnamefont {H.-Y.}\ \bibnamefont
  {Cheng}}\ and\ \bibinfo {author} {\bibfnamefont {C.-W.}\ \bibnamefont
  {Chiang}},\ }\href {\doibase 10.1103/PhysRevD.85.079903,
  10.1103/PhysRevD.85.034036} {\bibfield  {journal} {\bibinfo  {journal} {Phys.
  Rev.}\ }\textbf {\bibinfo {volume} {D85}},\ \bibinfo {pages} {034036}
  (\bibinfo {year} {2012})},\ \bibinfo {note} {[Erratum: Phys.
  Rev.D85,079903(2012)]},\ \Eprint {http://arxiv.org/abs/1201.0785}
  {arXiv:1201.0785 [hep-ph]} \BibitemShut {NoStop}%
\bibitem [{\citenamefont {Feldmann}\ \emph {et~al.}(2012)\citenamefont
  {Feldmann}, \citenamefont {Nandi},\ and\ \citenamefont
  {Soni}}]{Feldmann:2012js}%
  \BibitemOpen
  \bibfield  {author} {\bibinfo {author} {\bibfnamefont {T.}~\bibnamefont
  {Feldmann}}, \bibinfo {author} {\bibfnamefont {S.}~\bibnamefont {Nandi}}, \
  and\ \bibinfo {author} {\bibfnamefont {A.}~\bibnamefont {Soni}},\ }\href
  {\doibase 10.1007/JHEP06(2012)007} {\bibfield  {journal} {\bibinfo  {journal}
  {JHEP}\ }\textbf {\bibinfo {volume} {06}},\ \bibinfo {pages} {007} (\bibinfo
  {year} {2012})},\ \Eprint {http://arxiv.org/abs/1202.3795} {arXiv:1202.3795
  [hep-ph]} \BibitemShut {NoStop}%
\bibitem [{\citenamefont {Li}\ \emph {et~al.}(2012)\citenamefont {Li},
  \citenamefont {Lu},\ and\ \citenamefont {Yu}}]{Li:2012cfa}%
  \BibitemOpen
  \bibfield  {author} {\bibinfo {author} {\bibfnamefont {H.-n.}\ \bibnamefont
  {Li}}, \bibinfo {author} {\bibfnamefont {C.-D.}\ \bibnamefont {Lu}}, \ and\
  \bibinfo {author} {\bibfnamefont {F.-S.}\ \bibnamefont {Yu}},\ }\href
  {\doibase 10.1103/PhysRevD.86.036012} {\bibfield  {journal} {\bibinfo
  {journal} {Phys. Rev. D}\ }\textbf {\bibinfo {volume} {86}},\ \bibinfo
  {pages} {036012} (\bibinfo {year} {2012})},\ \Eprint
  {http://arxiv.org/abs/1203.3120} {arXiv:1203.3120 [hep-ph]} \BibitemShut
  {NoStop}%
\bibitem [{\citenamefont {Atwood}\ and\ \citenamefont
  {Soni}(2013)}]{Atwood:2012ac}%
  \BibitemOpen
  \bibfield  {author} {\bibinfo {author} {\bibfnamefont {D.}~\bibnamefont
  {Atwood}}\ and\ \bibinfo {author} {\bibfnamefont {A.}~\bibnamefont {Soni}},\
  }\href {\doibase 10.1093/ptep/ptt065} {\bibfield  {journal} {\bibinfo
  {journal} {PTEP}\ }\textbf {\bibinfo {volume} {2013}},\ \bibinfo {pages}
  {093B05} (\bibinfo {year} {2013})},\ \Eprint {http://arxiv.org/abs/1211.1026}
  {arXiv:1211.1026 [hep-ph]} \BibitemShut {NoStop}%
\bibitem [{\citenamefont {Grossman}\ and\ \citenamefont
  {Robinson}(2013)}]{Grossman:2012ry}%
  \BibitemOpen
  \bibfield  {author} {\bibinfo {author} {\bibfnamefont {Y.}~\bibnamefont
  {Grossman}}\ and\ \bibinfo {author} {\bibfnamefont {D.~J.}\ \bibnamefont
  {Robinson}},\ }\href {\doibase 10.1007/JHEP04(2013)067} {\bibfield  {journal}
  {\bibinfo  {journal} {JHEP}\ }\textbf {\bibinfo {volume} {04}},\ \bibinfo
  {pages} {067} (\bibinfo {year} {2013})},\ \Eprint
  {http://arxiv.org/abs/1211.3361} {arXiv:1211.3361 [hep-ph]} \BibitemShut
  {NoStop}%
\bibitem [{\citenamefont {Buccella}\ \emph {et~al.}(2019)\citenamefont
  {Buccella}, \citenamefont {Paul},\ and\ \citenamefont
  {Santorelli}}]{Buccella:2019kpn}%
  \BibitemOpen
  \bibfield  {author} {\bibinfo {author} {\bibfnamefont {F.}~\bibnamefont
  {Buccella}}, \bibinfo {author} {\bibfnamefont {A.}~\bibnamefont {Paul}}, \
  and\ \bibinfo {author} {\bibfnamefont {P.}~\bibnamefont {Santorelli}},\
  }\href {\doibase 10.1103/PhysRevD.99.113001} {\bibfield  {journal} {\bibinfo
  {journal} {Phys. Rev.}\ }\textbf {\bibinfo {volume} {D99}},\ \bibinfo {pages}
  {113001} (\bibinfo {year} {2019})},\ \Eprint
  {http://arxiv.org/abs/1902.05564} {arXiv:1902.05564 [hep-ph]} \BibitemShut
  {NoStop}%
\bibitem [{\citenamefont {Yu}\ \emph {et~al.}(2017)\citenamefont {Yu},
  \citenamefont {Wang},\ and\ \citenamefont {Li}}]{Yu:2017oky}%
  \BibitemOpen
  \bibfield  {author} {\bibinfo {author} {\bibfnamefont {F.-S.}\ \bibnamefont
  {Yu}}, \bibinfo {author} {\bibfnamefont {D.}~\bibnamefont {Wang}}, \ and\
  \bibinfo {author} {\bibfnamefont {H.-n.}\ \bibnamefont {Li}},\ }\href
  {\doibase 10.1103/PhysRevLett.119.181802} {\bibfield  {journal} {\bibinfo
  {journal} {Phys. Rev. Lett.}\ }\textbf {\bibinfo {volume} {119}},\ \bibinfo
  {pages} {181802} (\bibinfo {year} {2017})},\ \Eprint
  {http://arxiv.org/abs/1707.09297} {arXiv:1707.09297 [hep-ph]} \BibitemShut
  {NoStop}%
\bibitem [{\citenamefont {Brod}\ \emph
  {et~al.}(2012{\natexlab{b}})\citenamefont {Brod}, \citenamefont {Kagan},\
  and\ \citenamefont {Zupan}}]{Brod:2011re}%
  \BibitemOpen
  \bibfield  {author} {\bibinfo {author} {\bibfnamefont {J.}~\bibnamefont
  {Brod}}, \bibinfo {author} {\bibfnamefont {A.~L.}\ \bibnamefont {Kagan}}, \
  and\ \bibinfo {author} {\bibfnamefont {J.}~\bibnamefont {Zupan}},\ }\href
  {\doibase 10.1103/PhysRevD.86.014023} {\bibfield  {journal} {\bibinfo
  {journal} {Phys. Rev.}\ }\textbf {\bibinfo {volume} {D86}},\ \bibinfo {pages}
  {014023} (\bibinfo {year} {2012}{\natexlab{b}})},\ \Eprint
  {http://arxiv.org/abs/1111.5000} {arXiv:1111.5000 [hep-ph]} \BibitemShut
  {NoStop}%
\bibitem [{\citenamefont {Pirtskhalava}\ and\ \citenamefont
  {Uttayarat}(2012)}]{Pirtskhalava:2011va}%
  \BibitemOpen
  \bibfield  {author} {\bibinfo {author} {\bibfnamefont {D.}~\bibnamefont
  {Pirtskhalava}}\ and\ \bibinfo {author} {\bibfnamefont {P.}~\bibnamefont
  {Uttayarat}},\ }\href {\doibase 10.1016/j.physletb.2012.04.039} {\bibfield
  {journal} {\bibinfo  {journal} {Phys. Lett.}\ }\textbf {\bibinfo {volume}
  {B712}},\ \bibinfo {pages} {81} (\bibinfo {year} {2012})},\ \Eprint
  {http://arxiv.org/abs/1112.5451} {arXiv:1112.5451 [hep-ph]} \BibitemShut
  {NoStop}%
\bibitem [{\citenamefont {Johnson}\ and\ \citenamefont
  {Dudek}(2020)}]{Johnson:2020ilc}%
  \BibitemOpen
  \bibfield  {author} {\bibinfo {author} {\bibfnamefont {C.~T.}\ \bibnamefont
  {Johnson}}\ and\ \bibinfo {author} {\bibfnamefont {J.~J.}\ \bibnamefont
  {Dudek}},\ }\href@noop {} {\  (\bibinfo {year} {2020})},\ \Eprint
  {http://arxiv.org/abs/2012.00518} {arXiv:2012.00518 [hep-lat]} \BibitemShut
  {NoStop}%
\bibitem [{\citenamefont {Rendon}\ \emph {et~al.}(2020)\citenamefont {Rendon},
  \citenamefont {Leskovec}, \citenamefont {Meinel}, \citenamefont {Negele},
  \citenamefont {Paul}, \citenamefont {Petschlies}, \citenamefont {Pochinsky},
  \citenamefont {Silvi},\ and\ \citenamefont {Syritsyn}}]{Rendon:2020rtw}%
  \BibitemOpen
  \bibfield  {author} {\bibinfo {author} {\bibfnamefont {G.}~\bibnamefont
  {Rendon}}, \bibinfo {author} {\bibfnamefont {L.}~\bibnamefont {Leskovec}},
  \bibinfo {author} {\bibfnamefont {S.}~\bibnamefont {Meinel}}, \bibinfo
  {author} {\bibfnamefont {J.}~\bibnamefont {Negele}}, \bibinfo {author}
  {\bibfnamefont {S.}~\bibnamefont {Paul}}, \bibinfo {author} {\bibfnamefont
  {M.}~\bibnamefont {Petschlies}}, \bibinfo {author} {\bibfnamefont
  {A.}~\bibnamefont {Pochinsky}}, \bibinfo {author} {\bibfnamefont
  {G.}~\bibnamefont {Silvi}}, \ and\ \bibinfo {author} {\bibfnamefont
  {S.}~\bibnamefont {Syritsyn}},\ }\href {\doibase 10.1103/PhysRevD.102.114520}
  {\bibfield  {journal} {\bibinfo  {journal} {Phys. Rev.}\ }\textbf {\bibinfo
  {volume} {D102}},\ \bibinfo {pages} {114520} (\bibinfo {year} {2020})},\
  \Eprint {http://arxiv.org/abs/2006.14035} {arXiv:2006.14035 [hep-lat]}
  \BibitemShut {NoStop}%
\bibitem [{\citenamefont {Fischer}\ \emph {et~al.}(2020)\citenamefont
  {Fischer}, \citenamefont {Kostrzewa}, \citenamefont {Mai}, \citenamefont
  {Petschlies}, \citenamefont {Pittler}, \citenamefont {Ueding}, \citenamefont
  {Urbach},\ and\ \citenamefont {Werner}}]{Fischer:2020fvl}%
  \BibitemOpen
  \bibfield  {author} {\bibinfo {author} {\bibfnamefont {M.}~\bibnamefont
  {Fischer}}, \bibinfo {author} {\bibfnamefont {B.}~\bibnamefont {Kostrzewa}},
  \bibinfo {author} {\bibfnamefont {M.}~\bibnamefont {Mai}}, \bibinfo {author}
  {\bibfnamefont {M.}~\bibnamefont {Petschlies}}, \bibinfo {author}
  {\bibfnamefont {F.}~\bibnamefont {Pittler}}, \bibinfo {author} {\bibfnamefont
  {M.}~\bibnamefont {Ueding}}, \bibinfo {author} {\bibfnamefont
  {C.}~\bibnamefont {Urbach}}, \ and\ \bibinfo {author} {\bibfnamefont
  {M.}~\bibnamefont {Werner}} (\bibinfo {collaboration} {ETM}),\ }\href@noop {}
  {\  (\bibinfo {year} {2020})},\ \Eprint {http://arxiv.org/abs/2006.13805}
  {arXiv:2006.13805 [hep-lat]} \BibitemShut {NoStop}%
\bibitem [{\citenamefont {Bruno}\ and\ \citenamefont
  {Hansen}(2020)}]{Bruno:2020kyl}%
  \BibitemOpen
  \bibfield  {author} {\bibinfo {author} {\bibfnamefont {M.}~\bibnamefont
  {Bruno}}\ and\ \bibinfo {author} {\bibfnamefont {M.~T.}\ \bibnamefont
  {Hansen}},\ }\href@noop {} {\  (\bibinfo {year} {2020})},\ \Eprint
  {http://arxiv.org/abs/2012.11488} {arXiv:2012.11488 [hep-lat]} \BibitemShut
  {NoStop}%
\bibitem [{\citenamefont {Ablikim}\ \emph {et~al.}(2005)\citenamefont {Ablikim}
  \emph {et~al.}}]{Ablikim:2004wn}%
  \BibitemOpen
  \bibfield  {author} {\bibinfo {author} {\bibfnamefont {M.}~\bibnamefont
  {Ablikim}} \emph {et~al.} (\bibinfo {collaboration} {BES}),\ }\href {\doibase
  10.1016/j.physletb.2004.12.041} {\bibfield  {journal} {\bibinfo  {journal}
  {Phys. Lett.}\ }\textbf {\bibinfo {volume} {B607}},\ \bibinfo {pages} {243}
  (\bibinfo {year} {2005})},\ \Eprint {http://arxiv.org/abs/hep-ex/0411001}
  {arXiv:hep-ex/0411001 [hep-ex]} \BibitemShut {NoStop}%
\bibitem [{\citenamefont {Aaij}\ \emph
  {et~al.}(2014{\natexlab{b}})\citenamefont {Aaij} \emph
  {et~al.}}]{Aaij:2014emv}%
  \BibitemOpen
  \bibfield  {author} {\bibinfo {author} {\bibfnamefont {R.}~\bibnamefont
  {Aaij}} \emph {et~al.} (\bibinfo {collaboration} {LHCb}),\ }\href {\doibase
  10.1103/PhysRevD.89.092006} {\bibfield  {journal} {\bibinfo  {journal} {Phys.
  Rev.}\ }\textbf {\bibinfo {volume} {D89}},\ \bibinfo {pages} {092006}
  (\bibinfo {year} {2014}{\natexlab{b}})},\ \Eprint
  {http://arxiv.org/abs/1402.6248} {arXiv:1402.6248 [hep-ex]} \BibitemShut
  {NoStop}%
\bibitem [{\citenamefont {Dobbs}\ \emph {et~al.}(2015)\citenamefont {Dobbs},
  \citenamefont {Tomaradze}, \citenamefont {Xiao},\ and\ \citenamefont
  {Seth}}]{Dobbs:2015dwa}%
  \BibitemOpen
  \bibfield  {author} {\bibinfo {author} {\bibfnamefont {S.}~\bibnamefont
  {Dobbs}}, \bibinfo {author} {\bibfnamefont {A.}~\bibnamefont {Tomaradze}},
  \bibinfo {author} {\bibfnamefont {T.}~\bibnamefont {Xiao}}, \ and\ \bibinfo
  {author} {\bibfnamefont {K.~K.}\ \bibnamefont {Seth}},\ }\href {\doibase
  10.1103/PhysRevD.91.052006} {\bibfield  {journal} {\bibinfo  {journal} {Phys.
  Rev.}\ }\textbf {\bibinfo {volume} {D91}},\ \bibinfo {pages} {052006}
  (\bibinfo {year} {2015})},\ \Eprint {http://arxiv.org/abs/1502.01686}
  {arXiv:1502.01686 [hep-ex]} \BibitemShut {NoStop}%
\bibitem [{\citenamefont {Donoghue}\ and\ \citenamefont
  {Holstein}(1980)}]{Donoghue:1979fp}%
  \BibitemOpen
  \bibfield  {author} {\bibinfo {author} {\bibfnamefont {J.~F.}\ \bibnamefont
  {Donoghue}}\ and\ \bibinfo {author} {\bibfnamefont {B.~R.}\ \bibnamefont
  {Holstein}},\ }\href {\doibase 10.1103/PhysRevD.21.1334} {\bibfield
  {journal} {\bibinfo  {journal} {Phys. Rev.}\ }\textbf {\bibinfo {volume}
  {D21}},\ \bibinfo {pages} {1334} (\bibinfo {year} {1980})}\BibitemShut
  {NoStop}%
\bibitem [{\citenamefont {Lipkin}(1980)}]{Lipkin:1980es}%
  \BibitemOpen
  \bibfield  {author} {\bibinfo {author} {\bibfnamefont {H.~J.}\ \bibnamefont
  {Lipkin}},\ }\href {\doibase 10.1103/PhysRevLett.44.710} {\bibfield
  {journal} {\bibinfo  {journal} {Phys. Rev. Lett.}\ }\textbf {\bibinfo
  {volume} {44}},\ \bibinfo {pages} {710} (\bibinfo {year} {1980})}\BibitemShut
  {NoStop}%
\bibitem [{\citenamefont {Sorensen}(1981)}]{Sorensen:1981vu}%
  \BibitemOpen
  \bibfield  {author} {\bibinfo {author} {\bibfnamefont {C.}~\bibnamefont
  {Sorensen}},\ }\href {\doibase 10.1103/PhysRevD.23.2618} {\bibfield
  {journal} {\bibinfo  {journal} {Phys. Rev.}\ }\textbf {\bibinfo {volume}
  {D23}},\ \bibinfo {pages} {2618} (\bibinfo {year} {1981})}\BibitemShut
  {NoStop}%
\bibitem [{\citenamefont {Kamal}\ \emph {et~al.}(1988)\citenamefont {Kamal},
  \citenamefont {Sinha},\ and\ \citenamefont {Sinha}}]{Kamal:1988ub}%
  \BibitemOpen
  \bibfield  {author} {\bibinfo {author} {\bibfnamefont {A.~N.}\ \bibnamefont
  {Kamal}}, \bibinfo {author} {\bibfnamefont {N.}~\bibnamefont {Sinha}}, \ and\
  \bibinfo {author} {\bibfnamefont {R.}~\bibnamefont {Sinha}},\ }\href
  {\doibase 10.1007/BF01566918} {\bibfield  {journal} {\bibinfo  {journal} {Z.
  Phys.}\ }\textbf {\bibinfo {volume} {C41}},\ \bibinfo {pages} {207} (\bibinfo
  {year} {1988})}\BibitemShut {NoStop}%
\bibitem [{\citenamefont {Buccella}\ \emph {et~al.}(1990)\citenamefont
  {Buccella}, \citenamefont {Forte}, \citenamefont {Miele},\ and\ \citenamefont
  {Ricciardi}}]{Buccella:1990sp}%
  \BibitemOpen
  \bibfield  {author} {\bibinfo {author} {\bibfnamefont {F.}~\bibnamefont
  {Buccella}}, \bibinfo {author} {\bibfnamefont {M.}~\bibnamefont {Forte}},
  \bibinfo {author} {\bibfnamefont {G.}~\bibnamefont {Miele}}, \ and\ \bibinfo
  {author} {\bibfnamefont {G.}~\bibnamefont {Ricciardi}},\ }\href {\doibase
  10.1007/BF01565604} {\bibfield  {journal} {\bibinfo  {journal} {Z. Phys.}\
  }\textbf {\bibinfo {volume} {C48}},\ \bibinfo {pages} {47} (\bibinfo {year}
  {1990})}\BibitemShut {NoStop}%
\bibitem [{\citenamefont {Buccella}\ \emph {et~al.}(1992)\citenamefont
  {Buccella}, \citenamefont {Lusignoli}, \citenamefont {Miele},\ and\
  \citenamefont {Pugliese}}]{Buccella:1992ym}%
  \BibitemOpen
  \bibfield  {author} {\bibinfo {author} {\bibfnamefont {F.}~\bibnamefont
  {Buccella}}, \bibinfo {author} {\bibfnamefont {M.}~\bibnamefont {Lusignoli}},
  \bibinfo {author} {\bibfnamefont {G.}~\bibnamefont {Miele}}, \ and\ \bibinfo
  {author} {\bibfnamefont {A.}~\bibnamefont {Pugliese}},\ }\href {\doibase
  10.1007/BF01482585} {\bibfield  {journal} {\bibinfo  {journal} {Z. Phys.}\
  }\textbf {\bibinfo {volume} {C55}},\ \bibinfo {pages} {243} (\bibinfo {year}
  {1992})}\BibitemShut {NoStop}%
\bibitem [{\citenamefont {Buccella}\ \emph {et~al.}(1996)\citenamefont
  {Buccella}, \citenamefont {Lusignoli},\ and\ \citenamefont
  {Pugliese}}]{Buccella:1996uy}%
  \BibitemOpen
  \bibfield  {author} {\bibinfo {author} {\bibfnamefont {F.}~\bibnamefont
  {Buccella}}, \bibinfo {author} {\bibfnamefont {M.}~\bibnamefont {Lusignoli}},
  \ and\ \bibinfo {author} {\bibfnamefont {A.}~\bibnamefont {Pugliese}},\
  }\href {\doibase 10.1016/0370-2693(96)00460-1} {\bibfield  {journal}
  {\bibinfo  {journal} {Phys. Lett.}\ }\textbf {\bibinfo {volume} {B379}},\
  \bibinfo {pages} {249} (\bibinfo {year} {1996})},\ \Eprint
  {http://arxiv.org/abs/hep-ph/9601343} {arXiv:hep-ph/9601343 [hep-ph]}
  \BibitemShut {NoStop}%
\bibitem [{\citenamefont {Fajfer}\ and\ \citenamefont
  {Zupan}(1999)}]{Fajfer:1999hh}%
  \BibitemOpen
  \bibfield  {author} {\bibinfo {author} {\bibfnamefont {S.}~\bibnamefont
  {Fajfer}}\ and\ \bibinfo {author} {\bibfnamefont {J.}~\bibnamefont {Zupan}},\
  }\href {\doibase 10.1142/S0217751X99001950} {\bibfield  {journal} {\bibinfo
  {journal} {Int. J. Mod. Phys.}\ }\textbf {\bibinfo {volume} {A14}},\ \bibinfo
  {pages} {4161} (\bibinfo {year} {1999})},\ \Eprint
  {http://arxiv.org/abs/hep-ph/9903427} {arXiv:hep-ph/9903427 [hep-ph]}
  \BibitemShut {NoStop}%
\bibitem [{\citenamefont {Rosner}(1999)}]{Rosner:1999xd}%
  \BibitemOpen
  \bibfield  {author} {\bibinfo {author} {\bibfnamefont {J.~L.}\ \bibnamefont
  {Rosner}},\ }\href {\doibase 10.1103/PhysRevD.60.114026} {\bibfield
  {journal} {\bibinfo  {journal} {Phys. Rev.}\ }\textbf {\bibinfo {volume}
  {D60}},\ \bibinfo {pages} {114026} (\bibinfo {year} {1999})},\ \Eprint
  {http://arxiv.org/abs/hep-ph/9905366} {arXiv:hep-ph/9905366 [hep-ph]}
  \BibitemShut {NoStop}%
\bibitem [{\citenamefont {Gronau}(1999)}]{Gronau:1999zt}%
  \BibitemOpen
  \bibfield  {author} {\bibinfo {author} {\bibfnamefont {M.}~\bibnamefont
  {Gronau}},\ }\href {\doibase 10.1103/PhysRevLett.83.4005} {\bibfield
  {journal} {\bibinfo  {journal} {Phys. Rev. Lett.}\ }\textbf {\bibinfo
  {volume} {83}},\ \bibinfo {pages} {4005} (\bibinfo {year} {1999})},\ \Eprint
  {http://arxiv.org/abs/hep-ph/9908237} {arXiv:hep-ph/9908237} \BibitemShut
  {NoStop}%
\bibitem [{\citenamefont {Cheng}\ and\ \citenamefont
  {Chiang}(2010)}]{Cheng:2010ry}%
  \BibitemOpen
  \bibfield  {author} {\bibinfo {author} {\bibfnamefont {H.-Y.}\ \bibnamefont
  {Cheng}}\ and\ \bibinfo {author} {\bibfnamefont {C.-W.}\ \bibnamefont
  {Chiang}},\ }\href {\doibase 10.1103/PhysRevD.81.074021} {\bibfield
  {journal} {\bibinfo  {journal} {Phys. Rev.}\ }\textbf {\bibinfo {volume}
  {D81}},\ \bibinfo {pages} {074021} (\bibinfo {year} {2010})},\ \Eprint
  {http://arxiv.org/abs/1001.0987} {arXiv:1001.0987 [hep-ph]} \BibitemShut
  {NoStop}%
\bibitem [{\citenamefont {Fu-Sheng}\ \emph {et~al.}(2011)\citenamefont
  {Fu-Sheng}, \citenamefont {Wang},\ and\ \citenamefont {Lu}}]{Fusheng:2011tw}%
  \BibitemOpen
  \bibfield  {author} {\bibinfo {author} {\bibfnamefont {Y.}~\bibnamefont
  {Fu-Sheng}}, \bibinfo {author} {\bibfnamefont {X.-X.}\ \bibnamefont {Wang}},
  \ and\ \bibinfo {author} {\bibfnamefont {C.-D.}\ \bibnamefont {Lu}},\ }\href
  {\doibase 10.1103/PhysRevD.84.074019} {\bibfield  {journal} {\bibinfo
  {journal} {Phys. Rev.}\ }\textbf {\bibinfo {volume} {D84}},\ \bibinfo {pages}
  {074019} (\bibinfo {year} {2011})},\ \Eprint {http://arxiv.org/abs/1101.4714}
  {arXiv:1101.4714 [hep-ph]} \BibitemShut {NoStop}%
\bibitem [{\citenamefont {Biswas}\ \emph {et~al.}(2015)\citenamefont {Biswas},
  \citenamefont {Sinha},\ and\ \citenamefont {Abbas}}]{Biswas:2015aaa}%
  \BibitemOpen
  \bibfield  {author} {\bibinfo {author} {\bibfnamefont {A.}~\bibnamefont
  {Biswas}}, \bibinfo {author} {\bibfnamefont {N.}~\bibnamefont {Sinha}}, \
  and\ \bibinfo {author} {\bibfnamefont {G.}~\bibnamefont {Abbas}},\ }\href
  {\doibase 10.1103/PhysRevD.92.014032} {\bibfield  {journal} {\bibinfo
  {journal} {Phys. Rev.}\ }\textbf {\bibinfo {volume} {D92}},\ \bibinfo {pages}
  {014032} (\bibinfo {year} {2015})},\ \Eprint
  {http://arxiv.org/abs/1503.08176} {arXiv:1503.08176 [hep-ph]} \BibitemShut
  {NoStop}%
\bibitem [{\citenamefont {Gronau}\ and\ \citenamefont
  {Rosner}(2015)}]{Gronau:2015zga}%
  \BibitemOpen
  \bibfield  {author} {\bibinfo {author} {\bibfnamefont {M.}~\bibnamefont
  {Gronau}}\ and\ \bibinfo {author} {\bibfnamefont {J.~L.}\ \bibnamefont
  {Rosner}},\ }\href {\doibase 10.1103/PhysRevD.92.114018} {\bibfield
  {journal} {\bibinfo  {journal} {Phys. Rev.}\ }\textbf {\bibinfo {volume}
  {D92}},\ \bibinfo {pages} {114018} (\bibinfo {year} {2015})},\ \Eprint
  {http://arxiv.org/abs/1507.03565} {arXiv:1507.03565 [hep-ph]} \BibitemShut
  {NoStop}%
\bibitem [{\citenamefont {Eilam}\ \emph {et~al.}(1991)\citenamefont {Eilam},
  \citenamefont {Hewett},\ and\ \citenamefont {Soni}}]{Eilam:1991yv}%
  \BibitemOpen
  \bibfield  {author} {\bibinfo {author} {\bibfnamefont {G.}~\bibnamefont
  {Eilam}}, \bibinfo {author} {\bibfnamefont {J.~L.}\ \bibnamefont {Hewett}}, \
  and\ \bibinfo {author} {\bibfnamefont {A.}~\bibnamefont {Soni}},\ }\href
  {\doibase 10.1103/PhysRevLett.67.1979} {\bibfield  {journal} {\bibinfo
  {journal} {Phys. Rev. Lett.}\ }\textbf {\bibinfo {volume} {67}},\ \bibinfo
  {pages} {1979} (\bibinfo {year} {1991})}\BibitemShut {NoStop}%
\bibitem [{\citenamefont {Atwood}\ \emph {et~al.}(1995)\citenamefont {Atwood},
  \citenamefont {Eilam}, \citenamefont {Gronau},\ and\ \citenamefont
  {Soni}}]{Atwood:1994zm}%
  \BibitemOpen
  \bibfield  {author} {\bibinfo {author} {\bibfnamefont {D.}~\bibnamefont
  {Atwood}}, \bibinfo {author} {\bibfnamefont {G.}~\bibnamefont {Eilam}},
  \bibinfo {author} {\bibfnamefont {M.}~\bibnamefont {Gronau}}, \ and\ \bibinfo
  {author} {\bibfnamefont {A.}~\bibnamefont {Soni}},\ }\href {\doibase
  10.1016/0370-2693(94)01317-6, 10.1016/0370-2693(95)80017-R} {\bibfield
  {journal} {\bibinfo  {journal} {Phys. Lett.}\ }\textbf {\bibinfo {volume}
  {B341}},\ \bibinfo {pages} {372} (\bibinfo {year} {1995})},\ \Eprint
  {http://arxiv.org/abs/hep-ph/9409229} {arXiv:hep-ph/9409229 [hep-ph]}
  \BibitemShut {NoStop}%
\bibitem [{\citenamefont {Li}\ \emph {et~al.}(2014)\citenamefont {Li},
  \citenamefont {{L\"u}}, \citenamefont {Qin},\ and\ \citenamefont
  {Yu}}]{Li:2013xsa}%
  \BibitemOpen
  \bibfield  {author} {\bibinfo {author} {\bibfnamefont {H.-n.}\ \bibnamefont
  {Li}}, \bibinfo {author} {\bibfnamefont {C.-D.}\ \bibnamefont {{L\"u}}},
  \bibinfo {author} {\bibfnamefont {Q.}~\bibnamefont {Qin}}, \ and\ \bibinfo
  {author} {\bibfnamefont {F.-S.}\ \bibnamefont {Yu}},\ }\href {\doibase
  10.1103/PhysRevD.89.054006} {\bibfield  {journal} {\bibinfo  {journal} {Phys.
  Rev.}\ }\textbf {\bibinfo {volume} {D89}},\ \bibinfo {pages} {054006}
  (\bibinfo {year} {2014})},\ \Eprint {http://arxiv.org/abs/1305.7021}
  {arXiv:1305.7021 [hep-ph]} \BibitemShut {NoStop}%
\bibitem [{\citenamefont {Buccella}\ \emph {et~al.}(1993)\citenamefont
  {Buccella}, \citenamefont {Lusignoli}, \citenamefont {Mangano}, \citenamefont
  {Miele}, \citenamefont {Pugliese},\ and\ \citenamefont
  {Santorelli}}]{Buccella:1992sg}%
  \BibitemOpen
  \bibfield  {author} {\bibinfo {author} {\bibfnamefont {F.}~\bibnamefont
  {Buccella}}, \bibinfo {author} {\bibfnamefont {M.}~\bibnamefont {Lusignoli}},
  \bibinfo {author} {\bibfnamefont {G.}~\bibnamefont {Mangano}}, \bibinfo
  {author} {\bibfnamefont {G.}~\bibnamefont {Miele}}, \bibinfo {author}
  {\bibfnamefont {A.}~\bibnamefont {Pugliese}}, \ and\ \bibinfo {author}
  {\bibfnamefont {P.}~\bibnamefont {Santorelli}},\ }\href {\doibase
  10.1016/0370-2693(93)90402-4} {\bibfield  {journal} {\bibinfo  {journal}
  {Phys. Lett. B}\ }\textbf {\bibinfo {volume} {302}},\ \bibinfo {pages} {319}
  (\bibinfo {year} {1993})},\ \Eprint {http://arxiv.org/abs/hep-ph/9212253}
  {arXiv:hep-ph/9212253} \BibitemShut {NoStop}%
\bibitem [{\citenamefont {Dery}\ \emph {et~al.}(2021)\citenamefont {Dery},
  \citenamefont {Grossman}, \citenamefont {Schacht},\ and\ \citenamefont
  {Soffer}}]{Dery:2021mll}%
  \BibitemOpen
  \bibfield  {author} {\bibinfo {author} {\bibfnamefont {A.}~\bibnamefont
  {Dery}}, \bibinfo {author} {\bibfnamefont {Y.}~\bibnamefont {Grossman}},
  \bibinfo {author} {\bibfnamefont {S.}~\bibnamefont {Schacht}}, \ and\
  \bibinfo {author} {\bibfnamefont {A.}~\bibnamefont {Soffer}},\ }\href
  {\doibase 10.1007/JHEP05(2021)179} {\bibfield  {journal} {\bibinfo  {journal}
  {JHEP}\ }\textbf {\bibinfo {volume} {05}},\ \bibinfo {pages} {179} (\bibinfo
  {year} {2021})},\ \Eprint {http://arxiv.org/abs/2101.02560} {arXiv:2101.02560
  [hep-ph]} \BibitemShut {NoStop}%
\bibitem [{\citenamefont {Falk}\ \emph {et~al.}(1999)\citenamefont {Falk},
  \citenamefont {Nir},\ and\ \citenamefont {Petrov}}]{Falk:1999ts}%
  \BibitemOpen
  \bibfield  {author} {\bibinfo {author} {\bibfnamefont {A.~F.}\ \bibnamefont
  {Falk}}, \bibinfo {author} {\bibfnamefont {Y.}~\bibnamefont {Nir}}, \ and\
  \bibinfo {author} {\bibfnamefont {A.~A.}\ \bibnamefont {Petrov}},\ }\href
  {\doibase 10.1088/1126-6708/1999/12/019} {\bibfield  {journal} {\bibinfo
  {journal} {JHEP}\ }\textbf {\bibinfo {volume} {12}},\ \bibinfo {pages} {019}
  (\bibinfo {year} {1999})},\ \Eprint {http://arxiv.org/abs/hep-ph/9911369}
  {arXiv:hep-ph/9911369 [hep-ph]} \BibitemShut {NoStop}%
\bibitem [{\citenamefont {Golowich}\ and\ \citenamefont
  {Petrov}(1998)}]{Golowich:1998pz}%
  \BibitemOpen
  \bibfield  {author} {\bibinfo {author} {\bibfnamefont {E.}~\bibnamefont
  {Golowich}}\ and\ \bibinfo {author} {\bibfnamefont {A.~A.}\ \bibnamefont
  {Petrov}},\ }\href {\doibase 10.1016/S0370-2693(98)00329-3} {\bibfield
  {journal} {\bibinfo  {journal} {Phys. Lett.}\ }\textbf {\bibinfo {volume}
  {B427}},\ \bibinfo {pages} {172} (\bibinfo {year} {1998})},\ \Eprint
  {http://arxiv.org/abs/hep-ph/9802291} {arXiv:hep-ph/9802291 [hep-ph]}
  \BibitemShut {NoStop}%
\bibitem [{\citenamefont {Bergmann}\ \emph {et~al.}(2000)\citenamefont
  {Bergmann}, \citenamefont {Grossman}, \citenamefont {Ligeti}, \citenamefont
  {Nir},\ and\ \citenamefont {Petrov}}]{Bergmann:2000id}%
  \BibitemOpen
  \bibfield  {author} {\bibinfo {author} {\bibfnamefont {S.}~\bibnamefont
  {Bergmann}}, \bibinfo {author} {\bibfnamefont {Y.}~\bibnamefont {Grossman}},
  \bibinfo {author} {\bibfnamefont {Z.}~\bibnamefont {Ligeti}}, \bibinfo
  {author} {\bibfnamefont {Y.}~\bibnamefont {Nir}}, \ and\ \bibinfo {author}
  {\bibfnamefont {A.~A.}\ \bibnamefont {Petrov}},\ }\href {\doibase
  10.1016/S0370-2693(00)00772-3} {\bibfield  {journal} {\bibinfo  {journal}
  {Phys. Lett.}\ }\textbf {\bibinfo {volume} {B486}},\ \bibinfo {pages} {418}
  (\bibinfo {year} {2000})},\ \Eprint {http://arxiv.org/abs/hep-ph/0005181}
  {arXiv:hep-ph/0005181 [hep-ph]} \BibitemShut {NoStop}%
\bibitem [{\citenamefont {Falk}\ \emph {et~al.}(2002)\citenamefont {Falk},
  \citenamefont {Grossman}, \citenamefont {Ligeti},\ and\ \citenamefont
  {Petrov}}]{Falk:2001hx}%
  \BibitemOpen
  \bibfield  {author} {\bibinfo {author} {\bibfnamefont {A.~F.}\ \bibnamefont
  {Falk}}, \bibinfo {author} {\bibfnamefont {Y.}~\bibnamefont {Grossman}},
  \bibinfo {author} {\bibfnamefont {Z.}~\bibnamefont {Ligeti}}, \ and\ \bibinfo
  {author} {\bibfnamefont {A.~A.}\ \bibnamefont {Petrov}},\ }\href {\doibase
  10.1103/PhysRevD.65.054034} {\bibfield  {journal} {\bibinfo  {journal} {Phys.
  Rev.}\ }\textbf {\bibinfo {volume} {D65}},\ \bibinfo {pages} {054034}
  (\bibinfo {year} {2002})},\ \Eprint {http://arxiv.org/abs/hep-ph/0110317}
  {arXiv:hep-ph/0110317 [hep-ph]} \BibitemShut {NoStop}%
\bibitem [{\citenamefont {Falk}\ \emph {et~al.}(2004)\citenamefont {Falk},
  \citenamefont {Grossman}, \citenamefont {Ligeti}, \citenamefont {Nir},\ and\
  \citenamefont {Petrov}}]{Falk:2004wg}%
  \BibitemOpen
  \bibfield  {author} {\bibinfo {author} {\bibfnamefont {A.~F.}\ \bibnamefont
  {Falk}}, \bibinfo {author} {\bibfnamefont {Y.}~\bibnamefont {Grossman}},
  \bibinfo {author} {\bibfnamefont {Z.}~\bibnamefont {Ligeti}}, \bibinfo
  {author} {\bibfnamefont {Y.}~\bibnamefont {Nir}}, \ and\ \bibinfo {author}
  {\bibfnamefont {A.~A.}\ \bibnamefont {Petrov}},\ }\href {\doibase
  10.1103/PhysRevD.69.114021} {\bibfield  {journal} {\bibinfo  {journal} {Phys.
  Rev.}\ }\textbf {\bibinfo {volume} {D69}},\ \bibinfo {pages} {114021}
  (\bibinfo {year} {2004})},\ \Eprint {http://arxiv.org/abs/hep-ph/0402204}
  {arXiv:hep-ph/0402204 [hep-ph]} \BibitemShut {NoStop}%
\bibitem [{\citenamefont {Gerard}\ and\ \citenamefont
  {Hou}(1989)}]{Gerard:1988jj}%
  \BibitemOpen
  \bibfield  {author} {\bibinfo {author} {\bibfnamefont {J.-M.}\ \bibnamefont
  {Gerard}}\ and\ \bibinfo {author} {\bibfnamefont {W.-S.}\ \bibnamefont
  {Hou}},\ }\href {\doibase 10.1103/PhysRevLett.62.855} {\bibfield  {journal}
  {\bibinfo  {journal} {Phys. Rev. Lett.}\ }\textbf {\bibinfo {volume} {62}},\
  \bibinfo {pages} {855} (\bibinfo {year} {1989})}\BibitemShut {NoStop}%
\bibitem [{\citenamefont {Wolfenstein}(1991)}]{Wolfenstein:1990ks}%
  \BibitemOpen
  \bibfield  {author} {\bibinfo {author} {\bibfnamefont {L.}~\bibnamefont
  {Wolfenstein}},\ }\href {\doibase 10.1103/PhysRevD.43.151} {\bibfield
  {journal} {\bibinfo  {journal} {Phys. Rev. D}\ }\textbf {\bibinfo {volume}
  {43}},\ \bibinfo {pages} {151} (\bibinfo {year} {1991})}\BibitemShut
  {NoStop}%
\bibitem [{\citenamefont {Gerard}\ and\ \citenamefont
  {Hou}(1991)}]{Gerard:1990ni}%
  \BibitemOpen
  \bibfield  {author} {\bibinfo {author} {\bibfnamefont {J.-M.}\ \bibnamefont
  {Gerard}}\ and\ \bibinfo {author} {\bibfnamefont {W.-S.}\ \bibnamefont
  {Hou}},\ }\href {\doibase 10.1103/PhysRevD.43.2909} {\bibfield  {journal}
  {\bibinfo  {journal} {Phys. Rev. D}\ }\textbf {\bibinfo {volume} {43}},\
  \bibinfo {pages} {2909} (\bibinfo {year} {1991})}\BibitemShut {NoStop}%
\bibitem [{\citenamefont {Atwood}\ and\ \citenamefont
  {Soni}(1998)}]{Atwood:1997iw}%
  \BibitemOpen
  \bibfield  {author} {\bibinfo {author} {\bibfnamefont {D.}~\bibnamefont
  {Atwood}}\ and\ \bibinfo {author} {\bibfnamefont {A.}~\bibnamefont {Soni}},\
  }\href {\doibase 10.1103/PhysRevD.58.036005} {\bibfield  {journal} {\bibinfo
  {journal} {Phys. Rev. D}\ }\textbf {\bibinfo {volume} {58}},\ \bibinfo
  {pages} {036005} (\bibinfo {year} {1998})},\ \Eprint
  {http://arxiv.org/abs/hep-ph/9712287} {arXiv:hep-ph/9712287} \BibitemShut
  {NoStop}%
\bibitem [{\citenamefont {Atwood}\ \emph {et~al.}(2001)\citenamefont {Atwood},
  \citenamefont {Bar-Shalom}, \citenamefont {Eilam},\ and\ \citenamefont
  {Soni}}]{Atwood:2000tu}%
  \BibitemOpen
  \bibfield  {author} {\bibinfo {author} {\bibfnamefont {D.}~\bibnamefont
  {Atwood}}, \bibinfo {author} {\bibfnamefont {S.}~\bibnamefont {Bar-Shalom}},
  \bibinfo {author} {\bibfnamefont {G.}~\bibnamefont {Eilam}}, \ and\ \bibinfo
  {author} {\bibfnamefont {A.}~\bibnamefont {Soni}},\ }\href {\doibase
  10.1016/S0370-1573(00)00112-5} {\bibfield  {journal} {\bibinfo  {journal}
  {Phys. Rept.}\ }\textbf {\bibinfo {volume} {347}},\ \bibinfo {pages} {1}
  (\bibinfo {year} {2001})},\ \Eprint {http://arxiv.org/abs/hep-ph/0006032}
  {arXiv:hep-ph/0006032} \BibitemShut {NoStop}%
\bibitem [{\citenamefont {Zwicky}(2008)}]{Zwicky:2007vv}%
  \BibitemOpen
  \bibfield  {author} {\bibinfo {author} {\bibfnamefont {R.}~\bibnamefont
  {Zwicky}},\ }\href {\doibase 10.1103/PhysRevD.77.036004} {\bibfield
  {journal} {\bibinfo  {journal} {Phys. Rev. D}\ }\textbf {\bibinfo {volume}
  {77}},\ \bibinfo {pages} {036004} (\bibinfo {year} {2008})},\ \Eprint
  {http://arxiv.org/abs/0707.0677} {arXiv:0707.0677 [hep-ph]} \BibitemShut
  {NoStop}%
\bibitem [{\citenamefont {Bediaga}\ and\ \citenamefont
  {G\"obel}(2020)}]{Bediaga:2020qxg}%
  \BibitemOpen
  \bibfield  {author} {\bibinfo {author} {\bibfnamefont {I.}~\bibnamefont
  {Bediaga}}\ and\ \bibinfo {author} {\bibfnamefont {C.}~\bibnamefont
  {G\"obel}},\ }\href {\doibase 10.1016/j.ppnp.2020.103808} {\bibfield
  {journal} {\bibinfo  {journal} {Prog. Part. Nucl. Phys.}\ }\textbf {\bibinfo
  {volume} {114}},\ \bibinfo {pages} {103808} (\bibinfo {year} {2020})},\
  \Eprint {http://arxiv.org/abs/2009.07037} {arXiv:2009.07037 [hep-ex]}
  \BibitemShut {NoStop}%
\bibitem [{\citenamefont {{M\"uller}}\ \emph
  {et~al.}(2015{\natexlab{b}})\citenamefont {{M\"uller}}, \citenamefont
  {Nierste},\ and\ \citenamefont {Schacht}}]{Muller:2015lua}%
  \BibitemOpen
  \bibfield  {author} {\bibinfo {author} {\bibfnamefont {S.}~\bibnamefont
  {{M\"uller}}}, \bibinfo {author} {\bibfnamefont {U.}~\bibnamefont {Nierste}},
  \ and\ \bibinfo {author} {\bibfnamefont {S.}~\bibnamefont {Schacht}},\ }\href
  {\doibase 10.1103/PhysRevD.92.014004} {\bibfield  {journal} {\bibinfo
  {journal} {Phys. Rev.}\ }\textbf {\bibinfo {volume} {D92}},\ \bibinfo {pages}
  {014004} (\bibinfo {year} {2015}{\natexlab{b}})},\ \Eprint
  {http://arxiv.org/abs/1503.06759} {arXiv:1503.06759 [hep-ph]} \BibitemShut
  {NoStop}%
\bibitem [{\citenamefont {Grossman}\ \emph {et~al.}(2012)\citenamefont
  {Grossman}, \citenamefont {Kagan},\ and\ \citenamefont
  {Zupan}}]{Grossman:2012eb}%
  \BibitemOpen
  \bibfield  {author} {\bibinfo {author} {\bibfnamefont {Y.}~\bibnamefont
  {Grossman}}, \bibinfo {author} {\bibfnamefont {A.~L.}\ \bibnamefont {Kagan}},
  \ and\ \bibinfo {author} {\bibfnamefont {J.}~\bibnamefont {Zupan}},\ }\href
  {\doibase 10.1103/PhysRevD.85.114036} {\bibfield  {journal} {\bibinfo
  {journal} {Phys. Rev.}\ }\textbf {\bibinfo {volume} {D85}},\ \bibinfo {pages}
  {114036} (\bibinfo {year} {2012})},\ \Eprint {http://arxiv.org/abs/1204.3557}
  {arXiv:1204.3557 [hep-ph]} \BibitemShut {NoStop}%
\bibitem [{\citenamefont {Kwong}\ and\ \citenamefont
  {Rosen}(1993)}]{Kwong:1993ri}%
  \BibitemOpen
  \bibfield  {author} {\bibinfo {author} {\bibfnamefont {W.}~\bibnamefont
  {Kwong}}\ and\ \bibinfo {author} {\bibfnamefont {S.~P.}\ \bibnamefont
  {Rosen}},\ }\href {\doibase 10.1016/0370-2693(93)91843-C} {\bibfield
  {journal} {\bibinfo  {journal} {Phys. Lett. B}\ }\textbf {\bibinfo {volume}
  {298}},\ \bibinfo {pages} {413} (\bibinfo {year} {1993})}\BibitemShut
  {NoStop}%
\bibitem [{\citenamefont {Choi}\ \emph {et~al.}(1998)\citenamefont {Choi},
  \citenamefont {Lee},\ and\ \citenamefont {Song}}]{Choi:1998yx}%
  \BibitemOpen
  \bibfield  {author} {\bibinfo {author} {\bibfnamefont {S.~Y.}\ \bibnamefont
  {Choi}}, \bibinfo {author} {\bibfnamefont {J.}~\bibnamefont {Lee}}, \ and\
  \bibinfo {author} {\bibfnamefont {J.}~\bibnamefont {Song}},\ }\href {\doibase
  10.1016/S0370-2693(98)00872-7} {\bibfield  {journal} {\bibinfo  {journal}
  {Phys. Lett.}\ }\textbf {\bibinfo {volume} {B437}},\ \bibinfo {pages} {191}
  (\bibinfo {year} {1998})},\ \Eprint {http://arxiv.org/abs/hep-ph/9804268}
  {arXiv:hep-ph/9804268 [hep-ph]} \BibitemShut {NoStop}%
\bibitem [{\citenamefont {Cheng}(2003)}]{Cheng:2002wu}%
  \BibitemOpen
  \bibfield  {author} {\bibinfo {author} {\bibfnamefont {H.-Y.}\ \bibnamefont
  {Cheng}},\ }\href {\doibase 10.1140/epjc/s2002-01065-6} {\bibfield  {journal}
  {\bibinfo  {journal} {Eur. Phys. J.}\ }\textbf {\bibinfo {volume} {C26}},\
  \bibinfo {pages} {551} (\bibinfo {year} {2003})},\ \Eprint
  {http://arxiv.org/abs/hep-ph/0202254} {arXiv:hep-ph/0202254 [hep-ph]}
  \BibitemShut {NoStop}%
\bibitem [{\citenamefont {Colangelo}\ \emph {et~al.}(2010)\citenamefont
  {Colangelo}, \citenamefont {De~Fazio},\ and\ \citenamefont
  {Wang}}]{Colangelo:2010bg}%
  \BibitemOpen
  \bibfield  {author} {\bibinfo {author} {\bibfnamefont {P.}~\bibnamefont
  {Colangelo}}, \bibinfo {author} {\bibfnamefont {F.}~\bibnamefont {De~Fazio}},
  \ and\ \bibinfo {author} {\bibfnamefont {W.}~\bibnamefont {Wang}},\ }\href
  {\doibase 10.1103/PhysRevD.81.074001} {\bibfield  {journal} {\bibinfo
  {journal} {Phys. Rev. D}\ }\textbf {\bibinfo {volume} {81}},\ \bibinfo
  {pages} {074001} (\bibinfo {year} {2010})},\ \Eprint
  {http://arxiv.org/abs/1002.2880} {arXiv:1002.2880 [hep-ph]} \BibitemShut
  {NoStop}%
\bibitem [{\citenamefont {Ropertz}\ \emph {et~al.}(2018)\citenamefont
  {Ropertz}, \citenamefont {Hanhart},\ and\ \citenamefont
  {Kubis}}]{Ropertz:2018stk}%
  \BibitemOpen
  \bibfield  {author} {\bibinfo {author} {\bibfnamefont {S.}~\bibnamefont
  {Ropertz}}, \bibinfo {author} {\bibfnamefont {C.}~\bibnamefont {Hanhart}}, \
  and\ \bibinfo {author} {\bibfnamefont {B.}~\bibnamefont {Kubis}},\ }\href
  {\doibase 10.1140/epjc/s10052-018-6416-6} {\bibfield  {journal} {\bibinfo
  {journal} {Eur. Phys. J.}\ }\textbf {\bibinfo {volume} {C78}},\ \bibinfo
  {pages} {1000} (\bibinfo {year} {2018})},\ \Eprint
  {http://arxiv.org/abs/1809.06867} {arXiv:1809.06867 [hep-ph]} \BibitemShut
  {NoStop}%
\bibitem [{\citenamefont {Altmannshofer}\ \emph {et~al.}(2019)\citenamefont
  {Altmannshofer} \emph {et~al.}}]{Kou:2018nap}%
  \BibitemOpen
  \bibfield  {author} {\bibinfo {author} {\bibfnamefont {W.}~\bibnamefont
  {Altmannshofer}} \emph {et~al.} (\bibinfo {collaboration} {Belle-II}),\
  }\href {\doibase 10.1093/ptep/ptz106, 10.1093/ptep/ptaa008} {\bibfield
  {journal} {\bibinfo  {journal} {PTEP}\ }\textbf {\bibinfo {volume} {2019}},\
  \bibinfo {pages} {123C01} (\bibinfo {year} {2019})},\ \bibinfo {note}
  {[Erratum: PTEP2020,no.2,029201(2020)]},\ \Eprint
  {http://arxiv.org/abs/1808.10567} {arXiv:1808.10567 [hep-ex]} \BibitemShut
  {NoStop}%
\bibitem [{\citenamefont {Morningstar}\ and\ \citenamefont
  {Peardon}(1999)}]{MP99}%
  \BibitemOpen
  \bibfield  {author} {\bibinfo {author} {\bibfnamefont {C.~J.}\ \bibnamefont
  {Morningstar}}\ and\ \bibinfo {author} {\bibfnamefont {M.~J.}\ \bibnamefont
  {Peardon}},\ }\href {\doibase 10.1103/PhysRevD.60.034509} {\bibfield
  {journal} {\bibinfo  {journal} {Phys. Rev.}\ }\textbf {\bibinfo {volume}
  {D60}},\ \bibinfo {pages} {034509} (\bibinfo {year} {1999})},\ \Eprint
  {http://arxiv.org/abs/hep-lat/9901004} {arXiv:hep-lat/9901004 [hep-lat]}
  \BibitemShut {NoStop}%
\bibitem [{\citenamefont {Cornwall}\ and\ \citenamefont {Soni}(1984)}]{CS84}%
  \BibitemOpen
  \bibfield  {author} {\bibinfo {author} {\bibfnamefont {J.~M.}\ \bibnamefont
  {Cornwall}}\ and\ \bibinfo {author} {\bibfnamefont {A.}~\bibnamefont
  {Soni}},\ }\href {\doibase 10.1103/PhysRevD.29.1424} {\bibfield  {journal}
  {\bibinfo  {journal} {Phys. Rev.}\ }\textbf {\bibinfo {volume} {D29}},\
  \bibinfo {pages} {1424} (\bibinfo {year} {1984})}\BibitemShut {NoStop}%
\bibitem [{\citenamefont {Lees}\ \emph
  {et~al.}(2013{\natexlab{a}})\citenamefont {Lees} \emph
  {et~al.}}]{BaBar:2012bho}%
  \BibitemOpen
  \bibfield  {author} {\bibinfo {author} {\bibfnamefont {J.~P.}\ \bibnamefont
  {Lees}} \emph {et~al.} (\bibinfo {collaboration} {BaBar}),\ }\href {\doibase
  10.1103/PhysRevD.87.012004} {\bibfield  {journal} {\bibinfo  {journal} {Phys.
  Rev. D}\ }\textbf {\bibinfo {volume} {87}},\ \bibinfo {pages} {012004}
  (\bibinfo {year} {2013}{\natexlab{a}})},\ \Eprint
  {http://arxiv.org/abs/1209.3896} {arXiv:1209.3896 [hep-ex]} \BibitemShut
  {NoStop}%
\bibitem [{\citenamefont {Aaltonen}\ \emph {et~al.}(2014)\citenamefont
  {Aaltonen} \emph {et~al.}}]{CDF:2014wyb}%
  \BibitemOpen
  \bibfield  {author} {\bibinfo {author} {\bibfnamefont {T.~A.}\ \bibnamefont
  {Aaltonen}} \emph {et~al.} (\bibinfo {collaboration} {CDF}),\ }\href
  {\doibase 10.1103/PhysRevD.90.111103} {\bibfield  {journal} {\bibinfo
  {journal} {Phys. Rev. D}\ }\textbf {\bibinfo {volume} {90}},\ \bibinfo
  {pages} {111103} (\bibinfo {year} {2014})},\ \Eprint
  {http://arxiv.org/abs/1410.5435} {arXiv:1410.5435 [hep-ex]} \BibitemShut
  {NoStop}%
\bibitem [{\citenamefont {Aaij}\ \emph
  {et~al.}(2015{\natexlab{b}})\citenamefont {Aaij} \emph
  {et~al.}}]{LHCb:2015xyd}%
  \BibitemOpen
  \bibfield  {author} {\bibinfo {author} {\bibfnamefont {R.}~\bibnamefont
  {Aaij}} \emph {et~al.} (\bibinfo {collaboration} {LHCb}),\ }\href {\doibase
  10.1007/JHEP04(2015)043} {\bibfield  {journal} {\bibinfo  {journal} {JHEP}\
  }\textbf {\bibinfo {volume} {04}},\ \bibinfo {pages} {043} (\bibinfo {year}
  {2015}{\natexlab{b}})},\ \Eprint {http://arxiv.org/abs/1501.06777}
  {arXiv:1501.06777 [hep-ex]} \BibitemShut {NoStop}%
\bibitem [{\citenamefont {Stari\v{c}}\ \emph {et~al.}(2016)\citenamefont
  {Stari\v{c}} \emph {et~al.}}]{Belle:2015etc}%
  \BibitemOpen
  \bibfield  {author} {\bibinfo {author} {\bibfnamefont {M.}~\bibnamefont
  {Stari\v{c}}} \emph {et~al.} (\bibinfo {collaboration} {Belle}),\ }\href
  {\doibase 10.1016/j.physletb.2015.12.025} {\bibfield  {journal} {\bibinfo
  {journal} {Phys. Lett. B}\ }\textbf {\bibinfo {volume} {753}},\ \bibinfo
  {pages} {412} (\bibinfo {year} {2016})},\ \Eprint
  {http://arxiv.org/abs/1509.08266} {arXiv:1509.08266 [hep-ex]} \BibitemShut
  {NoStop}%
\bibitem [{\citenamefont {Aaij}\ \emph
  {et~al.}(2017{\natexlab{a}})\citenamefont {Aaij} \emph
  {et~al.}}]{LHCb:2017ejh}%
  \BibitemOpen
  \bibfield  {author} {\bibinfo {author} {\bibfnamefont {R.}~\bibnamefont
  {Aaij}} \emph {et~al.} (\bibinfo {collaboration} {LHCb}),\ }\href {\doibase
  10.1103/PhysRevLett.118.261803} {\bibfield  {journal} {\bibinfo  {journal}
  {Phys. Rev. Lett.}\ }\textbf {\bibinfo {volume} {118}},\ \bibinfo {pages}
  {261803} (\bibinfo {year} {2017}{\natexlab{a}})},\ \Eprint
  {http://arxiv.org/abs/1702.06490} {arXiv:1702.06490 [hep-ex]} \BibitemShut
  {NoStop}%
\bibitem [{\citenamefont {Aaij}\ \emph
  {et~al.}(2017{\natexlab{b}})\citenamefont {Aaij} \emph
  {et~al.}}]{LHCb:2016nxk}%
  \BibitemOpen
  \bibfield  {author} {\bibinfo {author} {\bibfnamefont {R.}~\bibnamefont
  {Aaij}} \emph {et~al.} (\bibinfo {collaboration} {LHCb}),\ }\href {\doibase
  10.1016/j.physletb.2017.01.061} {\bibfield  {journal} {\bibinfo  {journal}
  {Phys. Lett. B}\ }\textbf {\bibinfo {volume} {767}},\ \bibinfo {pages} {177}
  (\bibinfo {year} {2017}{\natexlab{b}})},\ \Eprint
  {http://arxiv.org/abs/1610.09476} {arXiv:1610.09476 [hep-ex]} \BibitemShut
  {NoStop}%
\bibitem [{\citenamefont {Staric}\ \emph {et~al.}(2008)\citenamefont {Staric}
  \emph {et~al.}}]{Belle:2008ddg}%
  \BibitemOpen
  \bibfield  {author} {\bibinfo {author} {\bibfnamefont {M.}~\bibnamefont
  {Staric}} \emph {et~al.} (\bibinfo {collaboration} {Belle}),\ }\href
  {\doibase 10.1016/j.physletb.2008.10.052} {\bibfield  {journal} {\bibinfo
  {journal} {Phys. Lett. B}\ }\textbf {\bibinfo {volume} {670}},\ \bibinfo
  {pages} {190} (\bibinfo {year} {2008})},\ \Eprint
  {http://arxiv.org/abs/0807.0148} {arXiv:0807.0148 [hep-ex]} \BibitemShut
  {NoStop}%
\bibitem [{\citenamefont {Csorna}\ \emph {et~al.}(2002)\citenamefont {Csorna}
  \emph {et~al.}}]{CLEO:2001lgl}%
  \BibitemOpen
  \bibfield  {author} {\bibinfo {author} {\bibfnamefont {S.~E.}\ \bibnamefont
  {Csorna}} \emph {et~al.} (\bibinfo {collaboration} {CLEO}),\ }\href {\doibase
  10.1103/PhysRevD.65.092001} {\bibfield  {journal} {\bibinfo  {journal} {Phys.
  Rev. D}\ }\textbf {\bibinfo {volume} {65}},\ \bibinfo {pages} {092001}
  (\bibinfo {year} {2002})},\ \Eprint {http://arxiv.org/abs/hep-ex/0111024}
  {arXiv:hep-ex/0111024} \BibitemShut {NoStop}%
\bibitem [{\citenamefont {Link}\ \emph {et~al.}(2000)\citenamefont {Link} \emph
  {et~al.}}]{FOCUS:2000ejh}%
  \BibitemOpen
  \bibfield  {author} {\bibinfo {author} {\bibfnamefont {J.~M.}\ \bibnamefont
  {Link}} \emph {et~al.} (\bibinfo {collaboration} {FOCUS}),\ }\href {\doibase
  10.1016/S0370-2693(00)01039-X} {\bibfield  {journal} {\bibinfo  {journal}
  {Phys. Lett. B}\ }\textbf {\bibinfo {volume} {491}},\ \bibinfo {pages} {232}
  (\bibinfo {year} {2000})},\ \bibinfo {note} {[Erratum: Phys.Lett.B 495,
  443--443 (2000)]},\ \Eprint {http://arxiv.org/abs/hep-ex/0005037}
  {arXiv:hep-ex/0005037} \BibitemShut {NoStop}%
\bibitem [{\citenamefont {Aitala}\ \emph {et~al.}(1998)\citenamefont {Aitala}
  \emph {et~al.}}]{E791:1997txw}%
  \BibitemOpen
  \bibfield  {author} {\bibinfo {author} {\bibfnamefont {E.~M.}\ \bibnamefont
  {Aitala}} \emph {et~al.} (\bibinfo {collaboration} {E791}),\ }\href {\doibase
  10.1016/S0370-2693(97)01570-0} {\bibfield  {journal} {\bibinfo  {journal}
  {Phys. Lett. B}\ }\textbf {\bibinfo {volume} {421}},\ \bibinfo {pages} {405}
  (\bibinfo {year} {1998})},\ \Eprint {http://arxiv.org/abs/hep-ex/9711003}
  {arXiv:hep-ex/9711003} \BibitemShut {NoStop}%
\bibitem [{\citenamefont {Nisar}\ \emph {et~al.}(2014)\citenamefont {Nisar}
  \emph {et~al.}}]{Belle:2014evd}%
  \BibitemOpen
  \bibfield  {author} {\bibinfo {author} {\bibfnamefont {N.~K.}\ \bibnamefont
  {Nisar}} \emph {et~al.} (\bibinfo {collaboration} {Belle}),\ }\href {\doibase
  10.1103/PhysRevLett.112.211601} {\bibfield  {journal} {\bibinfo  {journal}
  {Phys. Rev. Lett.}\ }\textbf {\bibinfo {volume} {112}},\ \bibinfo {pages}
  {211601} (\bibinfo {year} {2014})},\ \Eprint {http://arxiv.org/abs/1404.1266}
  {arXiv:1404.1266 [hep-ex]} \BibitemShut {NoStop}%
\bibitem [{\citenamefont {Aaij}\ \emph
  {et~al.}(2019{\natexlab{c}})\citenamefont {Aaij} \emph
  {et~al.}}]{LHCb:2019dwr}%
  \BibitemOpen
  \bibfield  {author} {\bibinfo {author} {\bibfnamefont {R.}~\bibnamefont
  {Aaij}} \emph {et~al.} (\bibinfo {collaboration} {LHCb}),\ }\href {\doibase
  10.1103/PhysRevLett.122.191803} {\bibfield  {journal} {\bibinfo  {journal}
  {Phys. Rev. Lett.}\ }\textbf {\bibinfo {volume} {122}},\ \bibinfo {pages}
  {191803} (\bibinfo {year} {2019}{\natexlab{c}})},\ \Eprint
  {http://arxiv.org/abs/1903.01150} {arXiv:1903.01150 [hep-ex]} \BibitemShut
  {NoStop}%
\bibitem [{\citenamefont {Lees}\ \emph
  {et~al.}(2013{\natexlab{b}})\citenamefont {Lees} \emph
  {et~al.}}]{BaBar:2012wep}%
  \BibitemOpen
  \bibfield  {author} {\bibinfo {author} {\bibfnamefont {J.~P.}\ \bibnamefont
  {Lees}} \emph {et~al.} (\bibinfo {collaboration} {BaBar}),\ }\href {\doibase
  10.1103/PhysRevD.87.052012} {\bibfield  {journal} {\bibinfo  {journal} {Phys.
  Rev. D}\ }\textbf {\bibinfo {volume} {87}},\ \bibinfo {pages} {052012}
  (\bibinfo {year} {2013}{\natexlab{b}})},\ \Eprint
  {http://arxiv.org/abs/1212.3003} {arXiv:1212.3003 [hep-ex]} \BibitemShut
  {NoStop}%
\bibitem [{\citenamefont {Ko}\ \emph {et~al.}(2013)\citenamefont {Ko} \emph
  {et~al.}}]{Belle:2012ygx}%
  \BibitemOpen
  \bibfield  {author} {\bibinfo {author} {\bibfnamefont {B.~R.}\ \bibnamefont
  {Ko}} \emph {et~al.} (\bibinfo {collaboration} {Belle}),\ }\href {\doibase
  10.1007/JHEP02(2013)098} {\bibfield  {journal} {\bibinfo  {journal} {JHEP}\
  }\textbf {\bibinfo {volume} {02}},\ \bibinfo {pages} {098} (\bibinfo {year}
  {2013})},\ \Eprint {http://arxiv.org/abs/1212.6112} {arXiv:1212.6112
  [hep-ex]} \BibitemShut {NoStop}%
\bibitem [{\citenamefont {Rubin}\ \emph {et~al.}(2006)\citenamefont {Rubin}
  \emph {et~al.}}]{CLEO:2005mti}%
  \BibitemOpen
  \bibfield  {author} {\bibinfo {author} {\bibfnamefont {P.}~\bibnamefont
  {Rubin}} \emph {et~al.} (\bibinfo {collaboration} {CLEO}),\ }\href {\doibase
  10.1103/PhysRevLett.96.081802} {\bibfield  {journal} {\bibinfo  {journal}
  {Phys. Rev. Lett.}\ }\textbf {\bibinfo {volume} {96}},\ \bibinfo {pages}
  {081802} (\bibinfo {year} {2006})},\ \Eprint
  {http://arxiv.org/abs/hep-ex/0512063} {arXiv:hep-ex/0512063} \BibitemShut
  {NoStop}%
\end{thebibliography}%
\bibliographystyle{apsrev4-1}

\end{document}